\documentclass{article}

\usepackage[final]{neurips_2025}

\usepackage{amsfonts}
\usepackage[utf8]{inputenc} 
\usepackage[T1]{fontenc}    
\usepackage{hyperref}       
\usepackage{url}            
\usepackage{booktabs}       
\usepackage{amsfonts}       
\usepackage{nicefrac}       
\usepackage{microtype}      
\usepackage{xcolor}         
\usepackage{makecell}
\usepackage[acronym]{glossaries}
\newacronym{MCS}{MCS}{Maximum Common Substructure}
\newacronym{MCES}{MCES}{Maximum Common Edge Subgraphs}
\newacronym{RASCAL}{RASCAL}{RApid Similarity CALculation}
\newacronym{PDB}{PDB}{Protein Data Bank}
\newacronym{RMSD}{RMSD}{Root Mean Square Deviation}
\newacronym{MD}{MD}{Molecular Dynamics}
\newacronym{FM}{FM}{Flow Matching}
\newacronym{GAFF2}{GAFF2}{General Amber Force Field 2}
\newacronym{FM-MAPO}{FMA-PO}{Flow Molecular Alignment with Pose Optimization}
\newacronym{FM-MA}{FMA}{Flow Molecular Alignment}
\newacronym{SB}{SB}{Structure-Based}
\newacronym{LB}{LB}{Ligand-Based}
\newacronym{PO}{PO}{Pose Optimization}
\newacronym{SMILES}{SMILES}{Simplified Molecular Input Line Entry System}
\newacronym{STS}{STS}{Shape-based Tanimoto Score}
\newacronym{PTS}{PTS}{Pharmacophore-based Tanimoto Score}
\newacronym{MHAwithEdgeBias}{MHAwithEdgeBias}{Multi-Head Attention with Edge Bias}
\newacronym{VFN}{VFN}{Vector Field Network}
\newacronym{ROCS}{ROCS}{Rapid Overlay of Chemical Structures}
\newacronym{PL}{PL}{protein-ligand}
\newacronym{MLP}{MLP}{multilayer perceptron}
\newacronym{HBD}{HBD}{Hydrogen Bond Donor}
\newacronym{HBA}{HBA}{Hydrogen Bond Acceptor}
\usepackage{geometry}
\usepackage{adjustbox}
\usepackage{graphicx}
\usepackage{subcaption}
\usepackage{multirow}
\usepackage{amsmath}
\usepackage{amssymb}
\usepackage{bbold}
\usepackage{float}
\usepackage[version=4]{mhchem}
\usepackage[ruled,vlined]{algorithm2e}

\title{Template-Guided 3D Molecular Pose Generation via Flow Matching and Differentiable Optimization}

\author{%
     Noémie Bergues$^{1}$
     \thanks{Equal contribution. Correspondence to: \texttt{\{noemie.bergues, arthur.carre\}@iktos.com}} \\
    \And
    Arthur Carré$^{1,2,3 \:}$\footnotemark[1] \\
    \And
    Paul Join-Lambert$^{1}$ \\
    \AND
    Brice Hoffmann$^{1}$ \\
    \And
    Arnaud Blondel$^{2}$ \\
    \And
    Hamza Tajmouati$^{1}$ \\
    \AND  
    \textnormal{$^1$ Iktos SA, 65 rue de Prony, 75017 Paris, France}\\
    \textnormal{$^2$ Institut Pasteur, Université Paris Cité, UMR 3528, Paris, France}\\
    \textnormal{$^3$ Doctoral school MTCI (ED 563), Paris Cité University, Paris, France}\\
}

\begin{document}

\maketitle

\begin{abstract}
Predicting the 3D conformation of small molecules within protein binding sites is a key challenge in drug design. When a crystallized reference ligand (template) is available, it provides geometric priors that can guide 3D pose prediction. We present a two-stage method for ligand conformation generation guided by such templates. In the first stage, we introduce a molecular alignment approach based on flow-matching to generate 3D coordinates for the ligand, using the template structure as a reference. In the second stage, a differentiable pose optimization procedure refines this conformation based on shape and pharmacophore similarities, internal energy, and, optionally, the protein binding pocket. We introduce a new benchmark of ligand pairs co-crystallized with the same target to evaluate our approach and show that it outperforms standard docking tools and open-access alignment methods, especially in cases involving low similarity to the template or high ligand flexibility.
\end{abstract}

\section{Introduction}
Drug design is a complex and resource-intensive process, with timelines that can span a decade and development costs estimated to be on the order of one billion dollars per approved drug \citep{wouters2020estimated}. To address these challenges, computational approaches have emerged to accelerate early-stage drug discovery and reduce associated costs. In particular, 3D-based methods show great promise by enabling \emph{in silico} screening and optimization of candidate molecules, thereby reducing reliance on expensive and time-consuming experimental assays. Two major 3D computational approaches are widely used: 3D \gls{LB} and 3D \gls{SB}. \gls{LB} methods leverage the 3D structure of known active compounds to identify or design structurally and functionally similar molecules \citep{acharya2011recent, petrovic2022virtual, bolcato2022value}, whereas \gls{SB} methods use the 3D structure of the target receptor to predict ligand binding modes or affinities \citep{anderson2003process}. Popular examples include \gls{LB} virtual screening tools such as \gls{ROCS} \citep{hawkins2007comparison}, and \gls{SB} molecular docking tools like AutoDock Vina \citep{trott2010autodock} and Glide \citep{friesner2004glide}.

Small molecule 3D alignment, or superposition, is an \gls{LB} approach that aims to spatially align a molecule with a known 3D template. A variety of methods have been developed for this task, reflecting the diverse ways in which molecular similarity can be defined and assessed. In their comprehensive review, Hönig \emph{et al}. \citep{honig2023small} categorize six major classes of alignment strategies: Gaussian volume overlap, field-based, graph-based, volume overlap optimization, distance-based, and shape-based methods. Despite these advances, accurately aligning molecules remains challenging, particularly when the template and query share limited structural or chemical features. Additionally, small molecule superposition is inherently an \gls{LB} approach \citep{honig2023small}, meaning it does not incorporate any structural information from the target protein, potentially overlooking critical receptor-ligand interactions that can drive binding. Given that many drug discovery projects benefit from the availability of experimentally resolved receptor-ligand complexes \citep{hubbard20053d, congreve2014structure}, there is an opportunity to develop hybrid methods that jointly leverage both ligand and receptor information to enhance the accuracy of molecular pose prediction.

This work introduces \gls{FM-MAPO}, a template-guided method for 3D molecular pose generation, which combines \gls{LB} alignment with structure-aware refinement. \gls{FM-MAPO} employs a \gls{FM} model conditioned on a 3D template molecule. The model generates 3D conformers for a query molecule from its 2D structure, aligning them spatially with the template. The poses are then refined via a coordinate-level differentiable optimization procedure integrating constraints based on shape alignment, pharmacophore similarity, binding pocket complementarity, and internal energy. To evaluate the performance of our method, we introduce AlignDockBench, a new benchmark comprising 369 \gls{PL} template–query pairs. Unlike existing cross-docking benchmarks such as the one proposed by FitDock \citep{yang_fitdock_2022}— which focuses on ligand pairs with high chemical similarity—AlignDockBench also includes pairs with lower similarity. This enables a more challenging and realistic assessment of \gls{LB} alignment and docking methods.

The main contributions of this work are as follows:
\begin{itemize}
    \item[-] A template-guided 3D molecular pose generation model with \gls{FM}, capable of producing accurate ligand conformations even in cases of low template similarity.
    \item[-] A pose refinement protocol operating directly on all atomic coordinates, improving pose accuracy and quality.
    \item[-] AlignDockBench, a benchmark for evaluating template-based docking accuracy.
\end{itemize}

\section{Related Work}
\paragraph{3D Alignment Methods.} 
In recent years, various advanced methods have been developed to improve the accuracy and efficiency of \gls{PL} docking and virtual screening, including approaches based on 3D molecular alignment. One of the earliest and most widely adopted tools in this area is \gls{ROCS}, which performs molecular alignment based on the overlap of Gaussian functions that represent molecular shape and, optionally, pharmacophoric features such as hydrogen bond donors, acceptors, and hydrophobic regions \citep{hawkins2007comparison}. More recently, LS-align \citep{hu_ls-align_2018} introduced a fast atom-level alignment algorithm that integrates interatomic distances, atomic mass, and chemical bond information, enabling both rigid and flexible alignments. Another notable method, FitDock \citep{yang_fitdock_2022}, improves docking accuracy by using template fitting to guide initial ligand conformations, followed by refinement with a scoring function derived from AutoDock Vina \citep{trott2010autodock}. ROSHAMBO \citep{atwi_roshambo_2024} is another recent method that combines shape and pharmacophore similarity using Gaussian volume overlap to perform 3D molecular alignment and similarity scoring.

\paragraph{Flow Matching for Biomolecular Applications.}
\gls{FM} is an emerging generative modeling framework that learns a continuous flow, mapping from a source distribution to a target distribution. Its adaptability has led to increasing adoption in biomolecular applications, including 3D conformer generation from 2D molecular graph \citep{hassan2025flow}. \gls{FM} has also been applied to molecular docking, with methods such as Harmonic Flow \citep{stark2024harmonic} and FlowDock \citep{morehead2024flowdock}, generating ligand conformations within protein binding pockets. AlphaFlow \citep{jing2024alphafold} extends \gls{FM} techniques to protein structure prediction, demonstrating their applicability to macromolecular modeling. Recently, \gls{FM} has been applied to \gls{MD} through MD-GEN \citep{jing2025generative}, a method that learns to generate physically realistic \gls{MD} trajectories. These advances underscore the growing potential of \gls{FM} to support a wide range of tasks in computational chemistry and structural biology.

\section{Method}\label{methods}
\subsection{Overview} 
\begin{figure}[h!]
  \centering
  \includegraphics[width=0.9\linewidth]{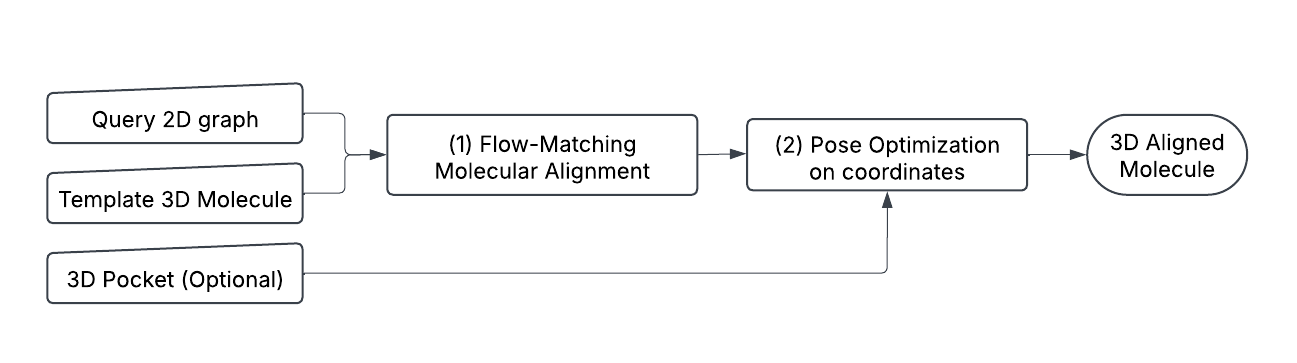}
  \caption{Overview of the \gls{FM-MAPO} pipeline. The method comprises two stages: (1) 3D \gls{FM-MA} of the ligand to the template; and (2) Pose Optimization, including shape and pharmacophores scores, energy minimization, and optional refinement based on protein pocket information.}
  \label{fig:overview_method}
\end{figure}

The present work introduces a Flow-Matching Molecular Alignment model followed by a Pose Optimization protocol (\gls{FM-MAPO}), a novel method for generating 3D molecular poses within a protein binding site from a 2D graph representation. \gls{FM-MAPO} uses a 3D reference ligand as a structural template (e.g., a crystallized ligand bound to the target protein), which serves as a spatial guide for predicting the pose of a query compound. The method consists of two main stages, as illustrated in Figure \ref{fig:overview_method}:  
\begin{enumerate}
\item \textbf{\gls{FM-MA}:} an initial 3D conformation of the ligand is generated using an \gls{FM} model trained to produce a conformer aligned with the reference ligand (template), given a 2D molecular graph as input.

\item \textbf{\gls{PO}:} the initial pose is refined through differentiable optimization on all coordinates using the following objectives:
\begin{itemize}
   \item[-] Shape and pharmacophore-based alignment optimization: the pose is optimized to maximize alignment with the reference ligand using shape and pharmacophore scoring functions. Pharmacophore is defined as a 3D arrangement of features (hydrogen bond donors/acceptors, hydrophobic groups, aromatic rings, etc.) for a molecule using the RDKit chemoinformatics library \citep{landrum_rdkit_2013}.
   \item[-] Internal energy optimization: the ligand's geometry is adjusted to minimize its internal (strain) energy, producing chemically valid conformations.
   \item[-] Protein pocket integration (optional): binding site information is incorporated to refine the pose, improve pharmacophore complementarity, and prevent steric clashes.
\end{itemize}
\end{enumerate}

\subsection{Flow-Matching Molecular Alignment}
\begin{figure}[h]
    \centering
    \begin{subfigure}[b]{0.24\textwidth}
        \centering
        \includegraphics[width=\textwidth]{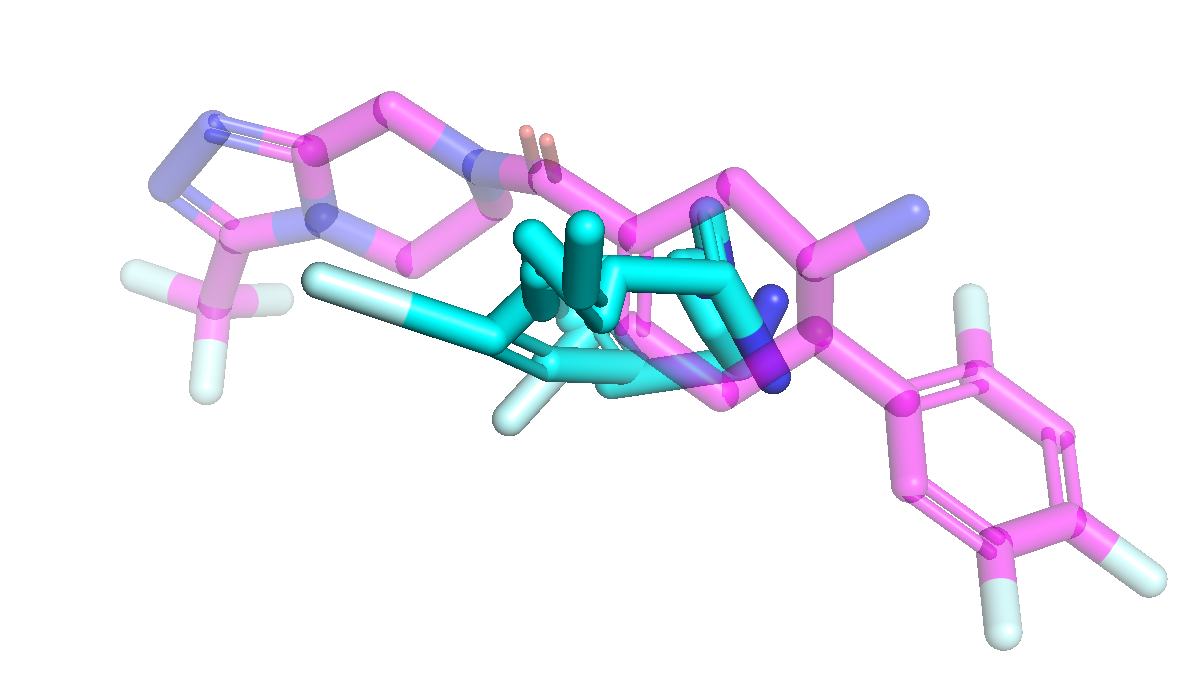}
        \caption{$t=0$}
    \end{subfigure}
    \hfill
    \begin{subfigure}[b]{0.24\textwidth}
        \centering
        \includegraphics[width=\textwidth]{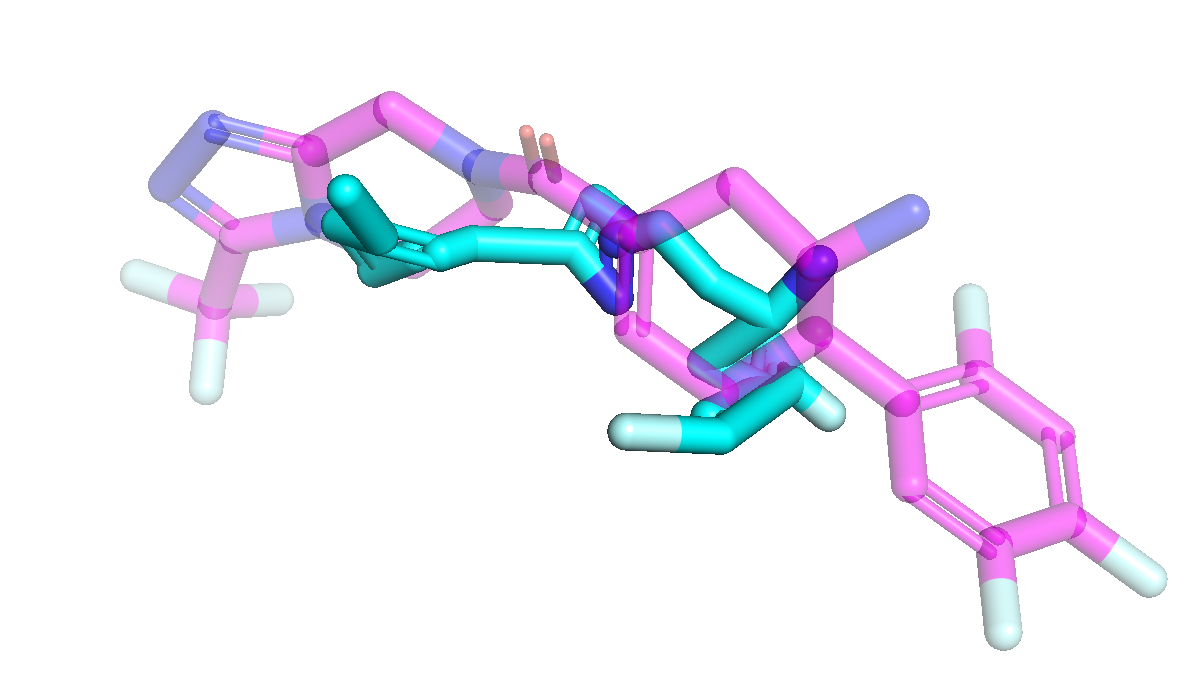}
        \caption{$t=0.5$}
    \end{subfigure}
    \hfill
    \begin{subfigure}[b]{0.24\textwidth}
        \centering
        \includegraphics[width=\textwidth]{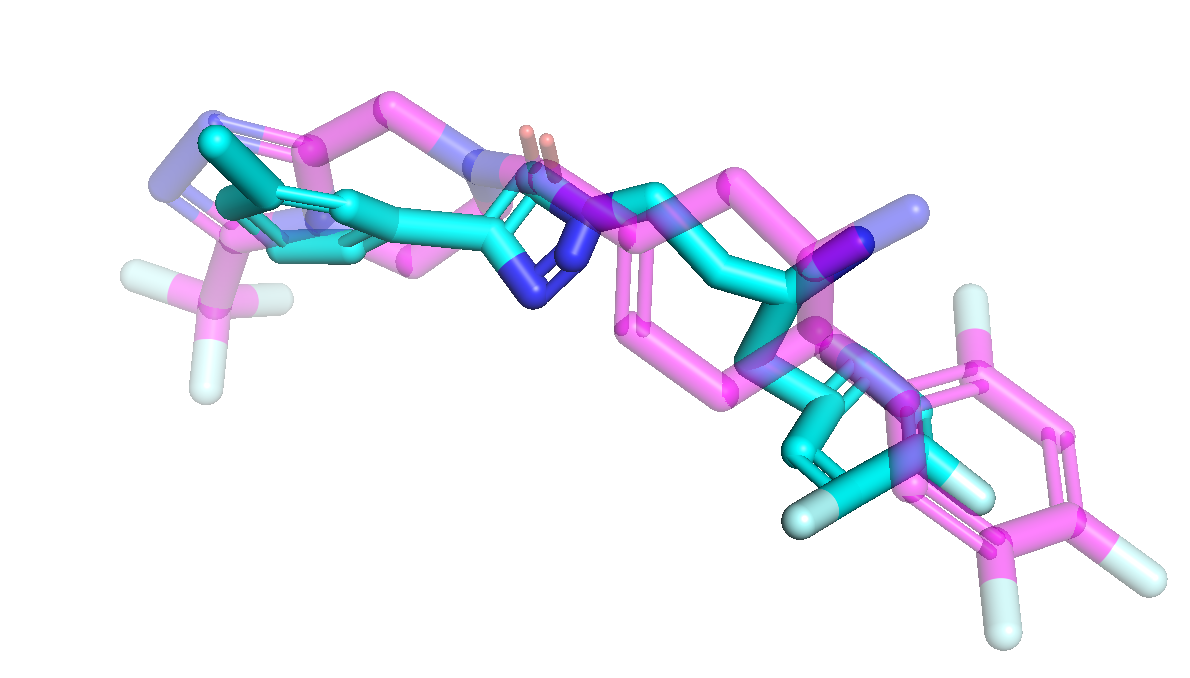}
        \caption{$t=0.75$}
    \end{subfigure}
    \hfill
    \begin{subfigure}[b]{0.24\textwidth}
        \centering
        \includegraphics[width=\textwidth]{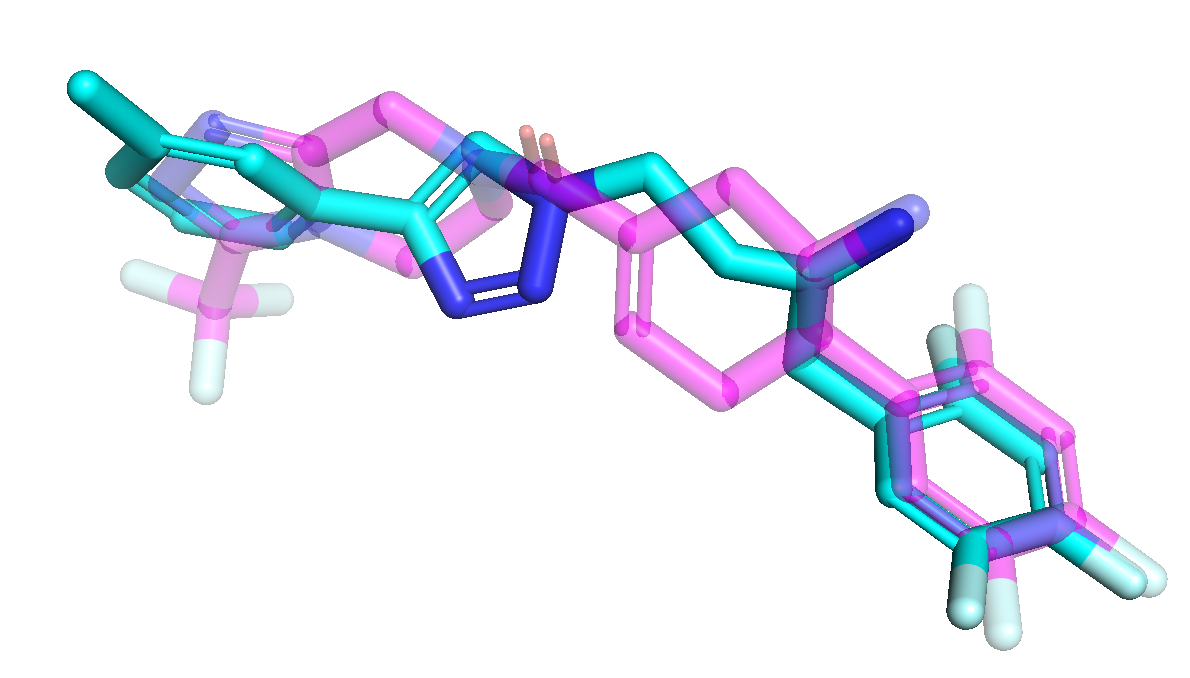}
        \caption{$t=1$}
    \end{subfigure}
    \caption{Evolution of the query ligand's conformation (in blue) during the FM-MA denoising process. The template ligand is shown in pink. Starting from a noisy initial conformation ($t=0$), the model progressively refines the 3D coordinates to align the query ligand with the template pose ($t=1$).}
    \label{fig:quatre_images}
\end{figure}

The \gls{FM-MA} module generates 3D conformations of a query ligand conditioned on the 3D structure of a reference ligand. Starting from a random 3D conformation of the query 2D graph, the model progressively denoises it to produce conformations that reproduce the ground-truth pose, spatially aligned with the reference ligand. This alignment process is illustrated in Figure \ref{fig:quatre_images}, which shows the evolution of the query conformation in blue, from initial noise ($t=0$) to final alignment with the reference structure in pink ($t=1$).
Inspired by pocket-guided generative models like DiffDock \citep{corso2022diffdock}, which condition ligand generation on protein pocket geometry, our method instead uses the 3D structure of a template ligand as the conditioning reference. An illustration of \gls{FM-MA} model pipeline is provided in Figure \ref{fig:FMA_schema}.

\paragraph{\gls{FM} model.}
\gls{FM} \citep{lipman2022flow} is a generative modeling approach that constructs a time-continuous mapping $\phi: [0,1] \times \mathbb{R}^d \to \mathbb{R}^d$ to transport a prior distribution $\rho_0$ to a target distribution $\rho_1$. This transformation is learned through a vector field (or velocity field) $v: [0,1] \times \mathbb{R}^d \to \mathbb{R}^d$ governing the sample evolution, described by the ordinary differential equation (ODE):
\begin{equation*}
\frac{d}{dt}\phi_t(x) = \mathbf{v}_t(\phi_t(x)), \quad \phi_0(x) = x.
\end{equation*}
That is, $\phi_t(x)$ represents the point $x$ transported along the vector field $v$ from time $0$ to time $t$. To construct a feasible transport trajectory, \citep{tong2023improving} \citep{albergo2023stochastic} introduces a time-differentiable interpolant:
\begin{equation*}
I_t: \mathbb{R}^d  \times \mathbb{R}^d \to \mathbb{R}^d \quad \text{such that} \quad I_0(x_0,x_1)=x_0 \quad \text{and} \quad I_1(x_0,x_1)=x_1.
\end{equation*}

Following \citep{hassan2025flow}, we choose a linear interpolation between samples $x_0 \sim \rho_0$ and $x_1 \sim \rho_1$:
\begin{equation*} I_t(\mathbf{x}_0, \mathbf{x}_1) = \alpha_t \mathbf{x}_1 + \beta_t \mathbf{x}_0, \quad \text{where} \quad \alpha_t = t, \quad \beta_t = 1 - t,
\end{equation*}
and similarly define the conditional probability path $\rho_t(\mathbf{x} | \mathbf{x}_0, \mathbf{x}_1)$  as a Gaussian distribution centered on $I_t(\mathbf{x}_0, \mathbf{x}_1)$ with variance $\sigma_t^2$: 
\begin{equation*} \rho_t(\mathbf{x} | \mathbf{x}_0, \mathbf{x}_1) = \mathcal{N}(x | I_t(\mathbf{x}_0, \mathbf{x}_1), \sigma_t^2 \mathbf{I}), \quad \text{where} \quad \sigma_t = \sigma \sqrt{t(1 - t)}. 
\end{equation*}

The time-dependent velocity field governing sample evolution is then given by:
\begin{equation*}
\mathbf{v}_t(\mathbf{x}) = \frac{d}{dt}\alpha_t \mathbf{x}_1 + \frac{d}{dt}\beta_t \mathbf{x}_0 + \frac{d}{dt}\sigma_t \mathbf{z}, \quad \mathbf{z} \sim \mathcal{N}(0, \mathbf{I}).
\end{equation*}
This results in the explicit formulation: $\mathbf{v}_t(\mathbf{x}) = \mathbf{x}_1 - \mathbf{x}_0 + \frac{1 - 2t}{2\sqrt{t(1 - t)}} \mathbf{z}$.

Given the probability path $\rho_t$ and the vector field $\mathbf{v}_t$, the \gls{FM} training objective is defined as:
\begin{equation*}
\mathcal{L} = \mathbb{E}_{t \sim \mathcal{U}(0,1), \mathbf{x} \sim \rho_t(\mathbf{x}_0, \mathbf{x}_1)} || \mathbf{v}_\theta(t, \mathbf{x}) - \mathbf{v}_t(\mathbf{x}) ||^2,
\end{equation*}
where $\mathbf{v}_\theta$ is the learned velocity field. The optimal solution ensures that $\mathbf{v}_\theta$ generates the probability path flow $\rho_t$ \citep{tong2023improving}.

For inference, a sample is drawn from $\rho_0$, and the final conformation $\mathbf{x}_1$ is obtained by numerically integrating the ODE using the Euler method and the learned velocity field $\mathbf{v}_\theta$:
\begin{equation*}
\mathbf{x}_{t+\Delta t} = \mathbf{x}_t + \mathbf{v}_\theta(t,  \mathbf{x}_t) \Delta t.
\end{equation*}

More details on the training and sampling algorithms are provided in supplementary \ref{training_inference}. 

\paragraph{Harmonic Prior.}  
Unlike diffusion-based models that typically require a Gaussian prior, the \gls{FM} framework allows flexibility in choosing the initial distribution $\rho_0$. To better capture the structural properties of molecular graphs, a harmonic prior is used, following the formulations in \citep{jing2023eigenfold} and \citep{stark2024harmonic}. This prior ensures that atom positions in the initial conformation reflect molecular connectivity by encouraging bonded atoms to remain close in sampled structures. The harmonic prior is defined as:
\begin{equation*}
    \rho_0(\mathbf{x}_0) \propto \exp\left(-\frac{1}{2} \mathbf{x}_0^T  \mathbf{L} \mathbf{x}_0 \right).
\end{equation*}
where $ \mathbf{L}$ is the laplacian of the molecular graph, and $x_0 \in \mathbb{R}^{1\times N}$ representing one of the three Cartesian coordinates (x, y, or z) of the $N$ atoms in the molecule. This formulation ensures that the sampled initial conformations maintain local structural coherence. The sampling algorithm is detailed in supplementary \ref{training_inference}.

\paragraph{Molecular Graph Construction.} 
\gls{FM-MA} is a molecular graph neural network that encodes a graph comprising both query and template ligands. It jointly represents their structures to predict the velocity field of the query ligand for conformation denoising. The graph representation is illustrated in Appendix \ref{graph_construction}, and is constructed as follows:

\begin{itemize}  
    \item[-] \textbf{Atomic-level nodes:} Each heavy atom in both ligands is represented as a node with features encoding atomic properties, including atomic number, total degree, formal charge, chiral tag, total number of hydrogen atoms, and hybridization state.
    \item[-] \textbf{Functional group nodes:} To capture higher-level chemical structures, ligands are decomposed into functional groups using the BRICS fragmentation method \citep{degen2008art}. Each resulting fragment is treated as a distinct node in the graph, with a scalar feature encoding the fragment size (number of atoms) assigned to each functional group node.
    \item[-] \textbf{Covalent edges:} Covalent bonds within each ligand are represented as graph edges, with associated edge features encoding bond type (e.g. single, double, triple, or aromatic), conjugation status, ring membership, and bond stereochemistry.
    \item[-] \textbf{Functional group connections:} Functional group nodes are connected to their corresponding constituent atoms, enabling feature aggregation from atomic-level representations. The edge features for these connections are scalar values set to zero.
    \item[-] \textbf{Dense edges:} Functional groups within and across the query and template ligands are fully connected. Distances between functional groups are encoded using a radial basis function (RBF), following \citep{unke2019physnet} and \citep{hassan2025flow}.
\end{itemize}  

\paragraph{Model Architecture.}
The constructed molecular graph, which includes both the query and template ligands, is first processed by a \gls{MLP} to embed node and edge features. Next, a \gls{MHAwithEdgeBias} encoder \citep{qiao2024state} is applied to extract topological and contextual information from the molecular graph. The attention mechanism facilitates effective information exchange between the two ligands, ensuring that the query ligand remains informed by the reference ligand. Finally, a time-dependent \gls{VFN}, inspired by \citep{qiao2024state}, predicts the velocity field of the query ligand, based on the molecular graph, the positions of the reference ligand, and positions of the query ligand at time $t$. Further architectural and algorithmic details are provided in Appendix \ref{ma_model}, and training details and hyperparameters in Appendix \ref{train}.

\subsection{Pose Optimization}
The initial pose of the query molecule, predicted by \gls{FM-MA}, is refined by directly optimizing its atomic coordinates. This process is formulated as a differentiable optimization problem, where gradient descent minimizes a weighted sum of scoring functions. These functions include ligand-based terms comparing the query ligand $\mathcal{M}_{\text{query}}$ to the reference ligand $\mathcal{M}_{\text{template}}$, and optionally, receptor-based terms involving the template pocket $\mathcal{P}_{\text{template}}$. Unlike traditional methods that optimize rigid-body transformations or torsion angles, \gls{PO} updates atomic positions directly, enabling flexible and fine-grained conformation refinement.

\paragraph{\gls{STS}.}
Each heavy atom in the molecular structure is modeled as a spherical Gaussian function, following the molecular volume representation introduced by \citet{grant1995gaussian, grant1996fast}. This formulation enables the computation of the molecular volume $V_{A}$, and the overlapping volume $V_{AB}$ of two molecules $A$ and $B$, as detailed in Appendix \ref{volume}. Tanimoto (or Jaccard) score based on atoms Gaussian volumes provides a normalized measure of shape similarity: 
\begin{equation*} 
Tanimoto_a(A,B) = \frac{V_{AB}^a}{V_A^a + V_B^a - V_{AB}^a}, 
\end{equation*} 
where $a$ refers to atom-based volumes. The \gls{STS} between the template $\mathcal{M}_{template}$ and the query $\mathcal{M}_{query}$ is defined as the $Tanimoto_{a}$ similarity between atoms:
\begin{equation*}
\mathcal{S}_{STS}(\mathcal{M}_{query}, \mathcal{M}_{ref})= Tanimoto_s(\mathcal{M}_{query}, \mathcal{M}_{ref}),
\end{equation*} 
and is differentiable with respect to the query ligand coordinates.

\paragraph{\gls{PTS}.}
Pharmacophoric features are represented using spherical Gaussian functions in a similar fashion. Each pharmacophoric feature type (e.g., hydrogen bond donor, acceptor, aromatic) is treated independently in the similarity computation. We define a pharmacophore-based alignment score that quantifies the degree of overlap between the pharmacophoric volumes of two ligands—higher values indicate that ligands share similar 3D pharmacophoric profile. This approach does not require explicit matching or pairing of features between the template and query molecules. Instead, the overlap between Gaussian spheres with the same pharmacophoric label is directly accumulated into the final pharmacophoric overlapping volume score. The \gls{PTS} computed on the pharmacophoric volumes between the query $\mathcal{M}_{query}$ and the reference $\mathcal{M}_{ref}$ is then defined as: 
\begin{equation*} 
\mathcal{S}_{PTS}(\mathcal{M}_{query}, \mathcal{M}_{template}) = Tanimoto_p(\mathcal{M}_{query}, \mathcal{M}_{template}),
\end{equation*} 
where $p$ refers to pharmacophore-based volumes with details provided in Appendix \ref{pharmaco}.

\paragraph{Protein Pocket Score (Optional).} Receptor-based terms are optionally included in optimization. Both atoms and pharmacophoric features of the template pocket $\mathcal{P}_{\text{template}}$ are modeled using Gaussian functions. Two scores are computed between the query ligand and the template pocket:
\begin{itemize}
    \item [-]  A shape clash penalty based on atomic overlap, using \gls{STS} to measure overlap between  $\mathcal{P}_{template}$ and $\mathcal{M}_{query}$.
    \item [-] A pharmacophoric complementarity, using \gls{PTS} to evaluate the alignment between the query ligand and template pocket pharmacophores. For pocket pharmacophore types, an inversion is applied to model complementary interactions. More details are provided in Appendix \ref{pharmaco}.
\end{itemize}

The resulting receptor-based pocket score is defined as: 
\begin{equation*}
\mathcal{S}_{pocket} =  \mathcal{S}_{PTS}(\mathcal{M}_{query}, \mathcal{P}_{template}) - \mathcal{S}_{STS}(\mathcal{M}_{query}, \mathcal{P}_{template}).
\end{equation*}
This formulation encourages pharmacophoric alignment while penalizing steric clashes between the ligand and the template binding pocket.

\paragraph{Internal Energy.}
Optimizing atomic coordinates directly, rather than restricting moves to translations, rotations, and torsions, offers greater flexibility but can easily lead to physically unrealistic molecular conformations. To ensure physically plausible structures, we incorporate the ligand's internal energy into the optimization process, thereby mitigating artifacts such as unrealistic bond lengths and angles that can arise from other scoring terms. The internal energy, $\mathcal{E}_{internal}$, is computed using molecular force fields, specifically \gls{GAFF2} \citep{wang2004gaff2}, which defines energy functions based on atomic types and spatial configurations. The detailed formulation is provided in Appendix \ref{energy}.

\paragraph{Optimization Objective.} The optimization objective combines all scores into a weighted loss function defined as follows:
\begin{align*}
    \mathcal{L}_{\text{optim}} = \: & -
    \alpha \: \mathcal{S}_{\text{STS}}(\mathcal{M}_{\text{query}}, \mathcal{M}_{\text{template}}) 
    - \beta \: \mathcal{S}_{\text{PTS}}(\mathcal{M}_{\text{query}}, \mathcal{M}_{\text{template}}) \\
    & - \omega \: \mathcal{S}_{\text{pocket}}(\mathcal{M}_{\text{query}}, \mathcal{P}_{\text{template}})+ \gamma \: \mathcal{E}_{\text{internal}}(\mathcal{M}_{\text{query}}).
\end{align*}
Further details on the optimization algorithm and hyperparameters are provided in Appendix \ref{PO}.

\subsection{Benchmark and training dataset construction} \label{datasets}
\paragraph{AlignDockBench.}\label{AlignDockBench} AlignDockBench is a curated benchmark designed to evaluate template-based molecular docking and 3D molecular alignment methods. Unlike common docking benchmarks such as PoseBusters \citep{buttenschoen2024posebusters}, which do not include reference ligands, AlignDockBench provides co-crystallized template ligands for each query. This enables the evaluation of methods that leverage known references to guide 3D conformation prediction. The benchmark includes $369$ \gls{PL} query structures, each associated with a corresponding \gls{PL} template structure. To construct AlignDockBench, we selected $61$ diverse \gls{PL} templates from the DUD-E \citep{mysinger_directory_2012} and DEKOIS \citep{bauer_evaluation_2013} datasets, spanning major target classes such as kinases, proteases, nuclear receptors, and GPCRs. For each template, we searched the \gls{PDB} for query \gls{PL} complexes whose binding pockets could be structurally aligned to the pocket template. Binding pockets were defined as all residues within $12$Å of the ligand. Using TM-align \citep{zhang_tm-align_2005}, we rigidly aligned (via translation and rotation) query pocket to the pocket template and retained only those with a backbone \gls{RMSD} below $1.2$Å. To ensure sufficient 3D structural similarity between ligands, we further filtered query-template pairs with an \gls{STS} score greater than $0.5$. The mapping of queries and associated templates is provided in Appendix \ref{benchmark_set}, along with \gls{PL} complex preparation details in Appendix \ref{structure_prep}. Protein class diversity is summarized in Figure \ref{fig:benchmark_pruned_diversity}. The benchmark is available at Zenodo \footnote{\url{https://anonymous.4open.science/r/AlignDockBench-6756/}}.

 \paragraph{Training set.} Training \gls{FM-MA} requires a dataset of molecular pairs with ground-truth alignments. Following the protocol used in AlignDockBench, we selected ligand pairs from the \gls{PDB} that bind to the same protein pocket, assuming that co-binding implies structural and functional similarity. We further restricted to pairs in which both ligands have molecular weights above 170 Da. Additionally, the training set was augmented with compounds from the ChEMBL database\footnote{\url{https://www.ebi.ac.uk/chembl/}}, following a protocol similar to that used for constructing the BindingNet and BindingNet2 datasets \citep{li2024high, zhu2025augmented}. To prevent train–test leakage, we excluded any training molecule with a Morgan fingerprint \citep{rogers2010extended} Tanimoto similarity greater than $0.5$ to any ligand in AlignDockBench (Figure \ref{fig:hist_tan_sim_morgan_fp_plot}). The resulting training set contains $301,348$ complex pairs covering $111,678$ unique molecules.

\section{Experiments} \label{experiments}
\paragraph{Experimental Setup.} We evaluate \gls{FM-MAPO} on AlignDockBench, comparing its performance to both traditional docking tools and state-of-the-art open-access alignment methods. Docking baselines include Vina (v1.2.5) \citep{trott2010autodock} and rDock (v24.04.204-legacy) \citep{ruiz2014rdock}, while alignment baselines include FitDock (v1.0.9) \citep{yang_fitdock_2022}, LS-align (Version J201704171741) \citep{hu_ls-align_2018}, and ROSHAMBO \citep{atwi_roshambo_2024}. Detailed configurations and hyperparameters used for each method are provided in Appendix \ref{experimental_set_up}. Alignment-based methods receive the 2D graph of the query molecule along with the 3D structure of the template ligand and, optionally, the protein. In contrast, in our experiments, docking methods do not use the template ligand information. For each method, we generate $10$ candidate poses and evaluate the top-ranked one using the scoring function originally designed for that method. The selected pose is then assessed by computing the \gls{RMSD} to the experimental ligand structure. Pose selection of our method is guided by a scoring function defined as:
\begin{equation*} \label{score}
\begin{aligned}
\mathcal{S}_{\text{score}} &= \mathcal{S}_{STS}(\mathcal{M}_{\text{query}}, \mathcal{M}_{\text{template}}) + \mathcal{S}_{PTS}(\mathcal{M}_{\text{query}}, \mathcal{M}_{\text{template}})  -\mathcal{S}_{\text{Vina}}(\mathcal{M}_{\text{query}}, \mathcal{P}_{template}),
\end{aligned}
\end{equation*}
where $\mathcal{S}_{\text{Vina}}$ represents the Vina docking score \citep{trott2010autodock}, rescaled to $[0, 1]$ interval using Min-Max normalization across the samples. Two pose selection strategies are explored: 

\begin{itemize}
    \item[-]\gls{FM-MAPO}: The top-ranked pose predicted by \gls{FM-MA}, according to $\mathcal{S}_{\text{score}}$, is selected and subsequently refined via \gls{PO}.
    \item[-]\gls{FM-MAPO}+: All sampled poses are first refined through \gls{PO}, and the final pose is selected according to $\mathcal{S}_{\text{score}}$.
\end{itemize}

\paragraph{Results.} Table \ref{tab:bench_results_crossdocking} summarizes the performance of each method in terms of mean \gls{RMSD}, the proportion of molecules with \gls{RMSD} below $2$Å, and the average runtime per molecule. Both \gls{FM-MAPO} and \gls{FM-MAPO}+ outperform competing methods, achieving the lowest mean \gls{RMSD} and the highest proportion of alignments with RMSD below $2$Å. Examples of poses generated with each method are provided in Figure \ref{fig:software_comparison_RMSDs}.

\begin{figure}
  \centering
  \begin{subfigure}[t]{0.48\linewidth}
    \includegraphics[width=\linewidth]{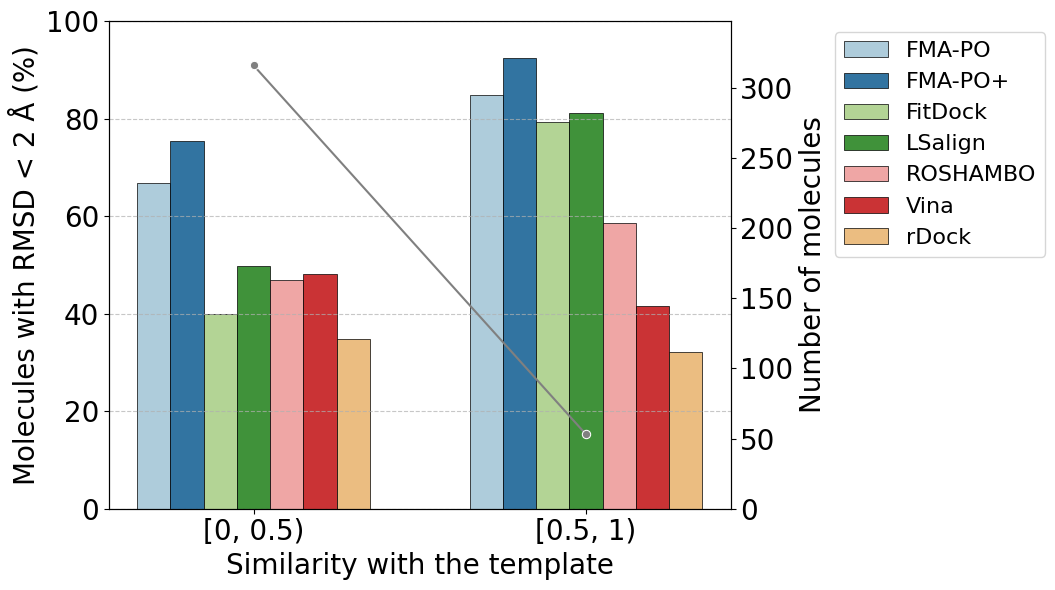}
    \caption{\scriptsize Stratified by the structure similarity between query and template.}
    \label{fig:all_results_by_similarity}
  \end{subfigure}
  \hfill
  \begin{subfigure}[t]{0.48\linewidth}
    \includegraphics[width=\linewidth]{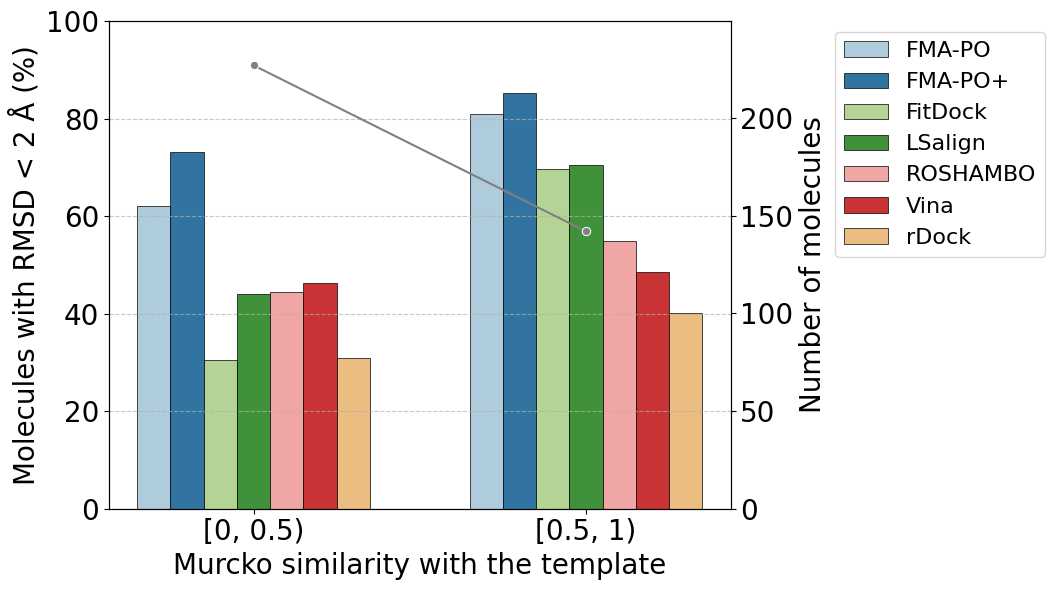}
    \caption{\scriptsize Stratified by the Murcko scaffold similarity between query and template.}
    \label{fig:all_results_by_murcko_similarity}
  \end{subfigure}

  \vspace{0.5em}  

  \begin{subfigure}[t]{0.48\linewidth}
    \includegraphics[width=\linewidth]{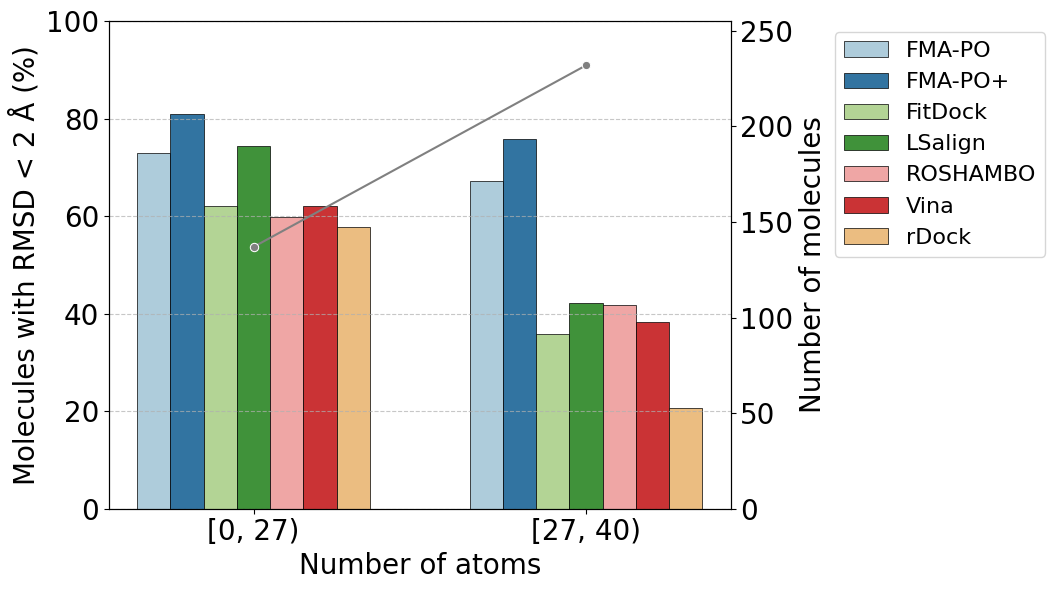}
    \caption{\scriptsize Stratified by the number of atoms in the query ligand.}
    \label{fig:all_results_by_na}
  \end{subfigure}
  \hfill
  \begin{subfigure}[t]{0.48\linewidth}
    \includegraphics[width=\linewidth]{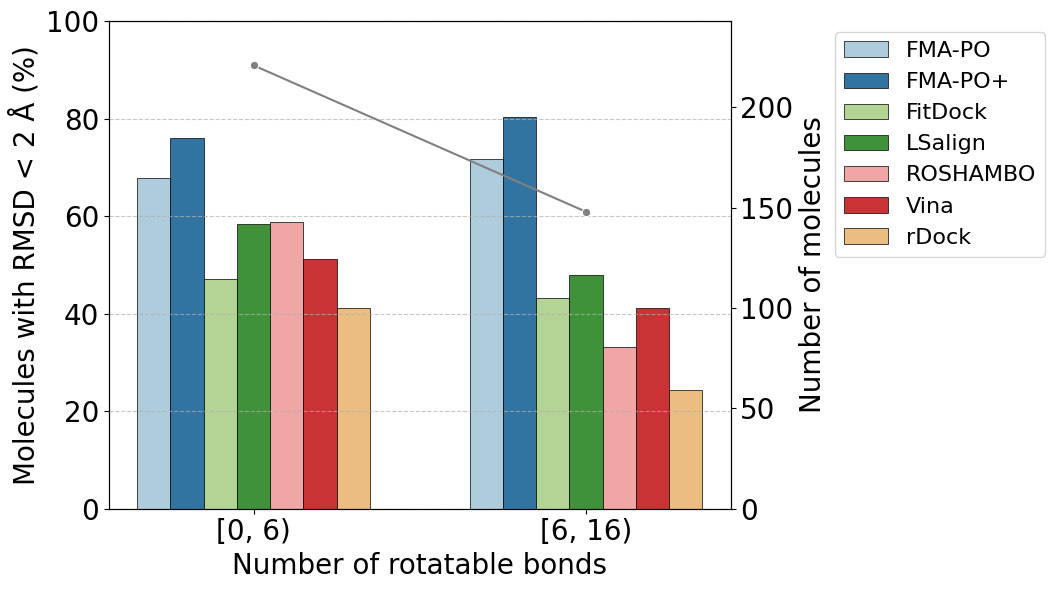}
    \caption{\scriptsize Stratified by the number of rotatable bonds in the query ligand.}
    \label{fig:all_results_by_rb}
  \end{subfigure}
    \vspace{0.7em} 
  \caption{Performance of molecular alignment and docking methods on AlignDockBench in a cross-docking scenario. Bar plots (left y-axis) indicate the percentage of molecules with RMSD below $2$Å, while the line plot (right y-axis) shows the number of molecules in each bin. Each figure stratifies the results based on a different structural or chemical property of the query ligand.}
  \label{fig:all_results_grid}
\end{figure}

To assess the effect of template similarity on alignment performance, AlignDockBench was divided into two similarity bins ($[0,0.5[$ and $[0.5,1.0]$), based on Tanimoto similarity between the query and template compounds. The similarity was computed using two approaches: first, with standard Morgan fingerprints \citep{rogers2010extended} derived from the complete molecular structures (Figure \ref{fig:all_results_by_similarity}); and second, with Morgan fingerprints computed on the molecules’ Generic Murcko scaffolds \citep{bemis1996properties}, which capture the core chemical framework by removing side chains and replacing all atoms with carbons (Figure \ref{fig:all_results_by_murcko_similarity}). We further analyzed how molecular complexity affects performance by evaluating results with respect to the number of atoms (Figure \ref{fig:all_results_by_na}) and the number of rotatable bonds (Figure \ref{fig:all_results_by_rb}) in the query molecules. For each bin and each method, we plotted the fraction of molecules achieving \gls{RMSD} below $2$Å (Figure \ref{fig:all_results_grid}).

As expected, alignment accuracy generally declines with decreasing molecular similarity to the template or increasing structural complexity. Nonetheless, \gls{FM-MAPO} and \gls{FM-MAPO}+ consistently outperform all baselines across all similarity and complexity bins, demonstrating their robustness in challenging scenarios. Indeed, \gls{FM-MA} generates poses without relying on explicit structural similarity, instead leveraging global spatial priors derived from the reference conformation. When the query molecule shares high similarity with the template, 3D \gls{LB} alignment methods such as LS-align and FitDock tend to outperform Vina and rDock \gls{SB} approaches, highlighting the value of exploiting reference ligand information to guide pose generation in these scenarios. 

All results were obtained under a cross-docking setup, where the pocket structure of the template ligand is used, simulating the realistic scenario in which the query ligand’s pocket structure is unavailable. Additional experiments were conducted under a redocking setup, using the query ligand’s own crystal pocket. Results are provided in the  Table \ref{tab:bench_results_redocking}. We further assessed the quality of generated poses using the PoseBusters test suite \citep{buttenschoen2024posebusters}, with detailed results available in Appendix \ref{PoseBuster_check}. Ablation studies are presented in Appendix \ref{ablation_study}. Comparison to Harmonic Flow \cite{stark2024harmonic} is provided in \ref{hf}.

\paragraph{Runtime.} In terms of computational efficiency, \gls{FM-MAPO} achieves a competitive average runtime of $3.66$ seconds per molecule. As expected, \gls{FM-MAPO}+ incurs a higher runtime ($27.96$ s per molecule) due to the additional \gls{PO} steps involving all samples but resulting in improved alignment accuracy. Both \gls{FM-MAPO} and \gls{FM-MAPO}+ benefit from GPU  \footnote{Experiments were run on a single NVIDIA T4 Tensor Core GPU.} acceleration during the \gls{FM-MA} stage, whereas conventional methods such as rDock and FitDock were executed using a single CPU core, and Vina was run with four CPU cores. More details are provided in \ref{discussion_runtime}.

\begin{table}
\caption{Performance comparison of 3D molecular alignment and docking methods on AlignDockBench in a crossdocking scenario. Methods marked with an (*) use GPU. For methods that did not align all 369 molecules, percentages are reported as X/Y, where the first value is calculated over aligned molecules only, and the second over the total set of 369 molecules.}
\label{tab:bench_results_crossdocking}
\vspace{0.7em} 
\begin{adjustbox}{max width=\textwidth}
\renewcommand{\arraystretch}{1.1} 
\centering
\begin{tabular}{lcccccc}
\toprule
\textbf{Method} & \makecell{\# of Molecules \\ Aligned} & \makecell{Mean RMSD \\ (Å ↓)} & \makecell{\% of Molecules \\ with RMSD \(<\) 2Å (↑)} & \makecell{Average \\ Runtime (s)} \\
\midrule
FMA$^*$  & 369/369 & 1.97 ± 1.36 &  64.77 & 0.83 \\
FMA-PO$^*$  & 369/369 & 1.86 ± 1.42 & 69.38 & 3.66 \\
FMA-PO$+^*$  & 369/369 & \textbf{1.62 ± 1.33} & \textbf{77.78} & 27.96 \\
FitDock & 313/369 & 2.93 ± 3.73 & 53.67 / 45.53 & 19.71\\
LSalign & 368/369 & 2.54 ± 2.00 & 54.35 / 54.2 & 5.67\\
ROSHAMBO$^*$ & 369/369 & 2.87 ± 2.09  & 30.35 & 4.90 \\
rDock & 369/369 & 4.52 ± 3.28  &  34.42 & 20.68 \\
Vina & 368/369 & 3.39 ± 2.81  & 47.28 / 47.15  & 72.98 \\
\bottomrule
\end{tabular}
\end{adjustbox}
\end{table}

\section{Discussion}\label{discussion}
Ligand alignment methods followed by 3D \gls{LB} similarity calculations are commonly used in virtual screening for hit identification \citep{jiang2021comprehensive}. \gls{FM-MAPO}+, combined with a scoring function that integrates \gls{STS}, \gls{PTS}, and optionally $\mathcal{S}_{pocket}$, fits naturally within this framework. Furthermore, the proposed method could be integrated into a \emph{de novo} generative design workflow as a reward signal, encouraging the generation of compounds that exhibit shape and pharmacophore similarity to a reference ligand, as well as complementarity to the protein binding pocket. Indeed, biasing molecular design towards compounds similar to known actives is particularly relevant during the hit discovery and hit-to-lead optimization stages \citep{bolcato2022value}.

\gls{FM-MAPO}+ is capable of generating 3D poses of candidate compounds that are directly suitable for downstream activity prediction tasks. Indeed, both deep learning-based scoring functions \citep{shen2023generalized, valsson2025narrowing} and physics-based approaches \citep{greenidge2013mm} rely on a 3D representation of the receptor-ligand complex. In addition, using \gls{FM-MAPO}+ generated alignments to augment training datasets represents a promising strategy to enhance the performance of predictive or generative models. This aligns with recent efforts such as BindingNet V2 \citep{zhu2025augmented}, which uses alignment-based data augmentation to boost the performance of deep learning models. In particular, \gls{FM-MA} itself could benefit from such augmented data to improve its performance in an iterative training process. Future work could explore conditioning \gls{FM-MA} on either a single or a set of reference ligands, along with the protein binding pocket, thereby incorporating more structural context into the pose generation process \cite{guan2025group}. Additionally, addressing flexible receptor scenarios, where both the 3D receptor and ligand are predicted, could be interesting. It is worth noting that, compared to non-deep learning methods, \gls{FM-MAPO} remains computationally more intensive, suggesting a potential direction for efficiency improvements.
\section{Conclusion}
This work presents a template-guided framework for 3D molecular pose generation that combines the strengths of \gls{FM} generative models with differentiable optimization. By conditioning the model on a reference ligand, the proposed \gls{FM-MAPO} approach achieves accurate molecular alignment, outperforming both classical docking tools and open-access alignment methods on the AlignDockBench template-based cross-docking benchmark. It remains robust even in challenging scenarios with low molecular similarity to the template and high ligand flexibility. These results highlight the potential of integrating known ligand geometries into generative frameworks to enhance pose prediction in structure-based drug design. Beyond performance gains, our framework enables sampling for data augmentation and direct integration with downstream predictive models. Future directions include incorporating receptor context into the generative process, extending the approach to flexible receptor scenarios (e.g., induced-fit), and assessing its impact on molecular design applications.

\section*{Acknowledgment}
We would like to express our sincere gratitude to the Iktos team for their exceptional
contributions to this project. The authors would like to thank IKTOS for supporting this study. We are also grateful to the reviewers for their valuable comments and suggestions, which have improved the manuscript. We further thank Ennys Gheyouche, Maoussi Lhuillier-Akakpo, and Juan Sanz García for their feedback and review of an early draft of this work.

\bibliographystyle{plainnat} 
\bibliography{references}

\appendix


\renewcommand{\thefigure}{S\arabic{figure}}
\renewcommand{\thetable}{S\arabic{table}}
\renewcommand{\thealgocf}{S\arabic{algocf}}
\renewcommand{\theequation}{S\arabic{equation}}

\setcounter{figure}{0}
\setcounter{table}{0}
\setcounter{algocf}{0}
\setcounter{equation}{0}
\setcounter{page}{1}

\section{Flow-Matching Molecular Alignment model details}
\begin{figure}[h!]
  \centering
  \includegraphics[height=0.50\linewidth]{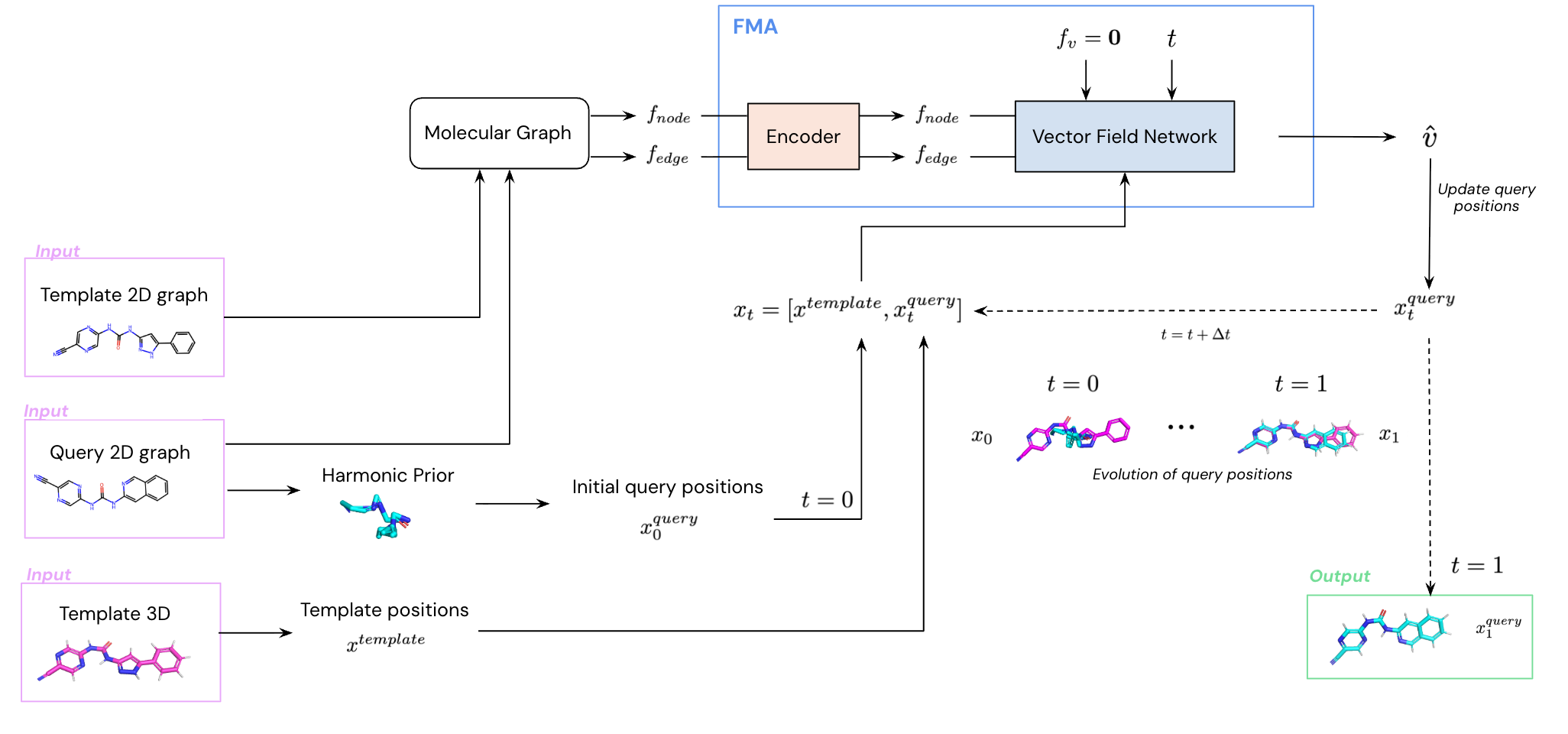}
  \caption{Illustration of FMA model pipeline.}
  \label{fig:FMA_schema}
\end{figure}

\subsection{Molecular graph construction}\label{graph_construction}
The \gls{FM-MA} model operates on a hybrid molecular graph $\mathcal{G}$ constructed from both the query and template ligands. The total number of atoms is denoted as $N_{\text{atoms}} = N_{\text{atoms}}^{\text{query}} + N_{\text{atoms}}^{\text{template}}$, where $N_{\text{atoms}}^{\text{query}}$ and $N_{\text{atoms}}^{\text{template}}$ are the numbers of atoms in the query and template molecules, respectively. The graph contains two types of nodes: atomic-level nodes and functional group nodes. Let $N_{\text{nodes}}$ denote the total number of nodes in the graph, including both atoms and functional groups. Edges include covalent bonds, dense inter-ligand functional group connections, and hierarchical links between atoms and their corresponding functional groups. The total number of edges is denoted by $N_{\text{edges}}$.

\begin{figure}[h!]
  \centering
  \includegraphics[height=0.55\linewidth]{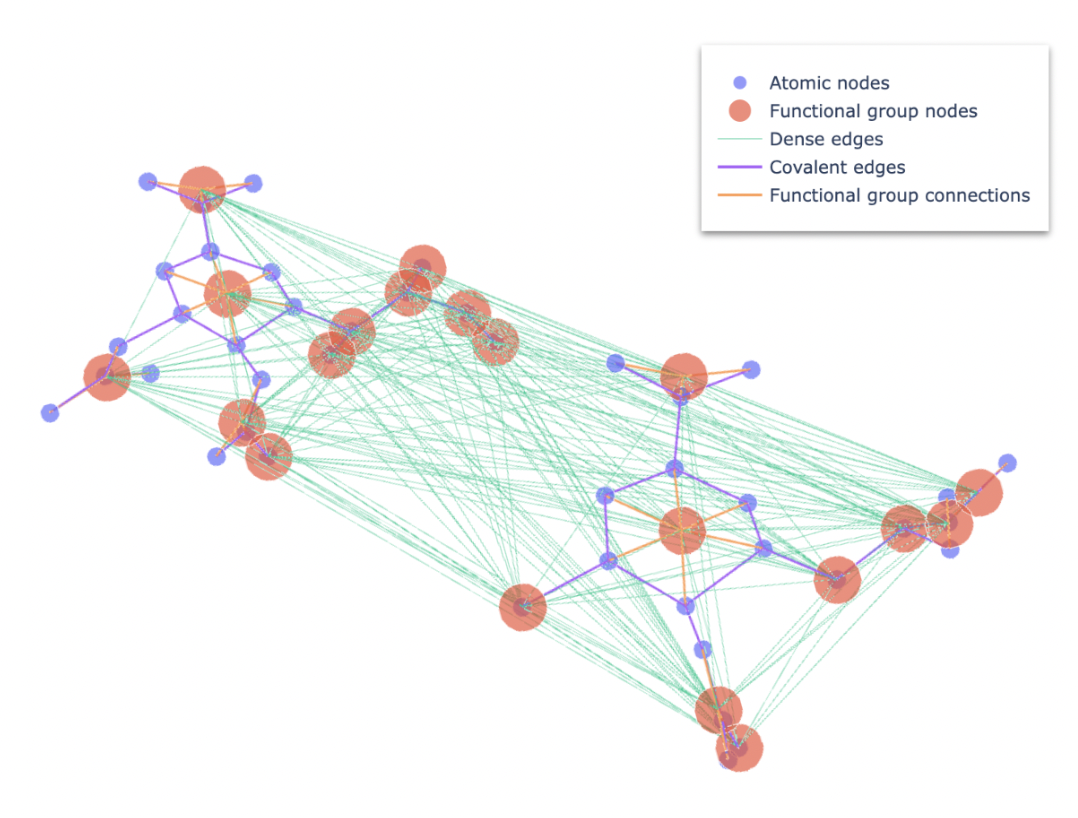}
  \caption{Molecular graph $\mathcal{G}$  with query and template ligand. Blue nodes represent atoms, while orange nodes denote functional groups. The graph incorporates multiple edge types: purple edges for covalent bonds, green edges for dense interconnections, and orange edges linking functional groups to their constituent atoms.}
  \label{fig:illustration_graph}
\end{figure}

\subsection{Flow-Matching Training and Inference Algorithms}\label{training_inference}
The inference and training procedures of the \gls{FM-MA} model are outlined in Algorithm \ref{algo:inference} and Algorithm \ref{algo:training}, respectively. In addition, the harmonic sampling process used to generate an initial random conformation from the reference structure is described in Algorithm \ref{algo:harmonic_sampling}.
\begin{algorithm}[H]
\caption{\textsc{Inference}}
\label{algo:inference}
\KwIn{Query 2D graph $G_{query}$, template molecule $\mathcal{M}_{template}$, number of sampling steps $N$,  number of samples $K$ }
$\mathcal{G} \gets$ Construct graph  with $G_{query}$ and $\mathcal{M}_{template}$ \\
\For{$i \gets 1$ to $K$}{
    Sample query positions $C_{i} \sim \textsc{HarmonicSampling}(G_{query})$ \\
    Center $C_{i}$ on $\mathcal{M}_{template}$ \\
    \For{$n \gets 0$ to $N-1$}{
    $t\gets \frac{n}{N}$ \\
    $\Delta t \gets \frac{1}{N}$ \\
    Predict vector field $\hat{v}  \gets \textsc{FMA}(\mathcal{G},t,C_i)$ \\
    Update positions $C_i \gets C_i+\hat{v} \times \Delta t $
    }
    } 
\Return $C_1, ..., C_K$
\end{algorithm}

\begin{algorithm}[h!]
\caption{\textsc{Training}}
\label{algo:training}
\KwIn{Query molecules with 2D graphs $[G_{query}^1, ...,G_{query}^K]$, true coordinates $[C^1,..., C^K]$ and associated template molecules $[\mathcal{M}_{template}^1,...,\mathcal{M}_{template}^K]$, number of epochs $N_{epochs}$, learning rate $\alpha$.}

\For{$n \gets 1$ to $N_{epochs}$}{
    \For{$i \gets 1$ to $K$}{
    Sample time $t\sim  \mathcal{U}([0,1])$ \\
    $\mathcal{G}^i \gets$ Construct graph  with $G_{query}^i$ and $\mathcal{M}_{template}^i$ \\
    Sample query positions $C_{0}^i \sim \textsc{HarmonicSampling}(G^i_{query})$ \\
    Center $C_0$ on $\mathcal{M}_{template}^i$ \\
    Sample $C_t^i = t \times C^i + (1-t) \times C_0^i + \sigma^2t(1-t) \mathbf{z}, \quad z \sim \mathcal{N}(0, \mathbf{I})$\\
    Compute vector field $u_t \gets C^i - C_0^i + \frac{1 - 2t}{2\sqrt{t(1 - t)}} \mathbf{z}$ \\
    Predict vector field $\hat{v}_{\theta}  \gets \textsc{FMA}(\mathcal{G}^i,t,C_t^i)$ \\
    Compute loss $\mathcal{L} \gets || \hat{v}_{\theta} - u_t ||^2$ \\
    Take gradient step $\theta \gets \theta - \alpha \times \Delta_{\theta}\mathcal{L}$
    }
    } 
\Return Trained ${v}_{\theta}$
\end{algorithm}

\begin{algorithm}[h!]
\caption{\textsc{HarmonicSampling}}
\label{algo:harmonic_sampling}
\KwIn{2D Molecular graph $G$}
$\mathbf{L} \gets$ Laplacian of $G$ \hfill \(\mathbf{L} \in  \mathbb{R}^{N_{atoms} \times N_{atoms}}  \) \\
$\mathbf{D},\mathbf{P} \gets$ Diagonalize Laplacian $\mathbf{L}$  \hfill \(\mathbf{D},\mathbf{P} \in  \mathbb{R}^{N_{atoms} \times N_{atoms}}  \) \\
Sample $x \sim \mathcal{N}(0,\mathbf{I})$ \hfill \( x \in  \mathbb{R}^{N_{atoms} \times 3}  \)  \\ 
$x \gets \mathbf{P}\mathbf{D}^{-\frac{1}{2}}x$ \\
\Return $x$
\end{algorithm}

\subsection{FMA Molecular Alignment model}\label{ma_model} 
\textsc{FMA} (Algorithm \ref{algo:MA}) takes as input a molecular graph $\mathcal{G}$ representing both query and template ligands, along with their atom coordinates. Positional features $x$ concatenates the ground-truth positions of the template atoms with the query atom coordinates at diffusion step $t$. The model outputs the vector field $v^{query}$ used for query pose prediction. Initially, node and edge features of $\mathcal{G}$ are independently embedded into vectors of dimensions $c_n$ and $c_e$, respectively, using separate MLPs. All node and edge embeddings are projected into a common feature space with identical dimensionality, thereby transforming the heterogeneous graph into a homogeneous representation. To preserve type-specific information, each node and edge is annotated with its corresponding type attribute. The resulting representation is processed by a stack of $N^{MHA}_\text{blocks}$ \textsc{MHAwithEdgeBias} layers (Algorithm S3 in \citep{qiao2024state}), jointly updating node and edge embeddings. The processed node embeddings are then passed to the $\textsc{VectorFieldNetwork}$ model (Algorithm \ref{algo:VFN}), which applies a stack of $N^{VFN}_\text{blocks}$ layers of $\textsc{PointSetAttentionwithEdgeBias}$ (Algorithm S8 in \citep{qiao2024state}), followed by a $\textsc{GatedUpdate}$ layer (Algorithm \ref{algo:GU}), adapted from \citep{qiao2024state}. These blocks operate directly on scalar and vector node representations, denoted $f_s$ and $f_v$ respectively, incorporating positional information $x$ and diffusion time $t$ to produce the final learned vector field $v^{query}$.

\begin{algorithm}[h!]
\caption{\textsc{FMA}}
\label{algo:MA}
\KwIn{Molecular graph $\mathcal{G}$ with adajency matrix $\mathbf{X}$, nodes features $f_{node}$, edges features $f_{edge}$, time $t$, template-query positions $x \in\mathbb{R}^{N_{atoms} \times 3} $ at time $t$, $N^{MHA}_\text{blocks},N^{MHA}_\text{heads}, N^{VFN}_\text{blocks}, N^{VFN}_\text{heads}$.}
\For{j in nodes types}{
        $f^j_{node} \gets\textsc{MLP}(f^j_{node})$  \hfill \(f^j_{node} \in  \mathbb{R}^{N^j_{nodes} \times c_n} \) 
    }
\For{j in edge types}{
        $f^j_{edge} \gets\textsc{MLP}(f^j_{edge})$ \hfill \(f^j_{edge} \in  \mathbb{R}^{N^j_{edges} \times c_e} \) 
    }
$\mathcal{G} \gets \textsc{ToHomogeneous}(\mathcal{G})$ \tcp*{Convert to Homogeneous Graph} 
\For{$k = 1$ to $N^{MHA}_\text{blocks}$}{
    $f_{node}', \_ ,z \gets \textsc{MHAwithEdgeBias}_k(f_{node}, f_{node}, f_{edge}, \mathbf{X},N^{MHA}_\text{heads})$ \hfill \(f_{node}' \in  \mathbb{R}^{N_{nodes} \times c_n} \) \\
    $f_{node} \gets \textsc{MLP}_k(f_{node}+f_{node}')+f_{node}$ \hfill \(f_{node} \in  \mathbb{R}^{N_{nodes} \times c_n} \) \\
    $f_{edge}\gets \textsc{MLP}_k(f_{edge}+\textsc{LinearNoBias}_k(z))+f_{edge} $ \hfill \(f_{edge} \in  \mathbb{R}^{N_{edges} \times c_n} \)

}
$ f_v \gets \mathbf{0} \in  \mathbb{R}^{N_{nodes} \times 3  \times c_n} $ \tcp*{Initialize vector features as zero tensors}
$ x \gets x - \text{mean}(x) $ \tcp*{Center positions}
$ v \gets \textsc{VectorFieldNetwork}(f_{node}, f_v, f_{edge}, $x$ , t, N^{VFN}_\text{blocks}, N^{VFN}_\text{heads})$ \hfill \(v \in  \mathbb{R}^{N_{nodes} \times 3} \)  \\
$v \gets$ Extract atomic nodes from $v$  \hfill \(v \in  \mathbb{R}^{N_{atoms} \times 3} \)  \\
$v^{query} \gets$ Extract Query vector from $v$ \hfill \(v^{query} \in  \mathbb{R}^{N^{query} _{atoms} \times 3} \) \\
\Return $v^{query}$
\end{algorithm}

\begin{algorithm}[h!]
\caption{\textsc{VectorFieldNetwork}}
\label{algo:VFN}
\KwIn{Scalar features $fs\in  \mathbb{R}^{N_{nodes}\times c_n}$, vector features $fv\in  \mathbb{R}^{N_{nodes} \times 3 \times c_n}$, edges features $fe\in  \mathbb{R}^{N_{edges}\times c_e} $, query-template positions x, time t,  $N_\text{blocks}$, $N_\text{heads}$}
$fs \leftarrow \text{\textsc{concat}}(fs, t)$ \hfill \(fs \in  \mathbb{R}^{N_{nodes} \times (c_n+1)} \) \\
\For{$k = 1$ to $N_\text{blocks}$}{
        $fs, fv \leftarrow \text{\textsc{PointSetAttentionwithEdgeBias}}_k(fs, fv, fe, x, N_\text{heads})$ \hfill \(fs \in  \mathbb{R}^{N_{nodes} \times (c_n+1)}, fv \in  \mathbb{R}^{N_{nodes}\times 3 \times (c_n+1)}  \)\\
        $fs, fv \leftarrow \text{\textsc{GatedUpdate}}_k(fs, fv, (c_n+1))$\hfill \(fs \in  \mathbb{R}^{N_{nodes} \times (c_n+1)}, fv \in  \mathbb{R}^{N_{nodes}\times 3 \times (c_n+1)}  \)
    }
 \textunderscore $ \: , v \leftarrow \text{\textsc{GatedUpdate}}(fs, fv,1)$ \hfill \(v \in  \mathbb{R}^{N_{nodes} \times 3} \)  \\
\Return $v$ \\
\end{algorithm}

\begin{algorithm}[h!]
\caption{\textsc{GatedUpdate}}
\label{algo:GU}
\KwIn{Scalar features $fs\in  \mathbb{R}^{N_{nodes}\times c_n}$, vector features $fv\in  \mathbb{R}^{N_{nodes} \times 3 \times c_n}$, output dimension $d$  }
$f_{loc} \leftarrow \text{\textsc{concat}}(fs, \lVert fv \lVert_2)$  \hfill \(f_{loc} \in  \mathbb{R}^{N_{nodes}  \times (c_n+c_n)} \)\\
$f_{loc} \leftarrow \text{\textsc{LayerNormMLP}}(f_{loc})$ \hfill \(f_{loc} \in  \mathbb{R}^{N_{nodes}  \times (c_n+d)} \) \\
$fs, f_{gate} \leftarrow \text{\textsc{split}}(f_{loc})$ \hfill \(fs \in  \mathbb{R}^{N_{nodes}  \times c},   f_{gate} \in  \mathbb{R}^{N_{nodes}  \times 1 \times d } \) \\
$f_{gate} \leftarrow \text{\textsc{Sigmoid}}(f_{gate})$\\
$fv \leftarrow \text{\textsc{LinearNoBias}}(fv)$ \hfill \(fv \in  \mathbb{R}^{N_{nodes} \times 3 \times d} \) \\
$fv \leftarrow fv \odot f_{gate}$  \tcp*{Element wise multiplication}
\Return $fv$
\end{algorithm}

\subsection{Training Details and Hyperparameters}\label{train}
We trained \gls{FM-MA} for 100 epochs with a batch size of 128 on a single NVIDIA GeForce RTX 2080 Ti GPU. We used the AdamW optimizer with a learning rate of $3 \times 10^{-4}$ and a weight decay of $10^{-5}$. Hyperparameters are detailed in Table \ref{tab:hyperparameters}.

\begin{table}[h!]
\caption{Hyperparameters used for the training of \textsc{FMA}.}
\label{tab:hyperparameters}
\vspace{0.7em}
\centering
\begin{adjustbox}{max width=\textwidth}
\renewcommand{\arraystretch}{1.1}
\begin{tabular}{l c}
\toprule
\textbf{Hyperparameter} &  \\
\midrule
\vspace{0.5em}
Nodes embedding dimension ($c_n$) & 128 \\
\vspace{0.5em}
Edges embedding dimension ($c_e$) & 32 \\
\vspace{0.5em}
\textsc{MHAwithEdgeBias} heads ($N^{\text{MHA}}_{\text{heads}}$) & 6 \\
\vspace{0.5em}
\textsc{MHAwithEdgeBias} blocks ($N^{\text{MHA}}_{\text{blocks}}$) & 6 \\
\vspace{0.5em}
\textsc{VectorFieldNetwork} heads ($N^{\text{VFN}}_{\text{heads}}$) & 6 \\
\vspace{0.5em}
\textsc{VectorFieldNetwork} blocks ($N^{\text{VFN}}_{\text{blocks}}$) & 6 \\
\vspace{0.5em}
Noise scale ($\sigma$) & 0.1 \\
\midrule
Total parameters & 3.5M \\
\bottomrule
\end{tabular}
\end{adjustbox}
\end{table}

\section{Pose Optimization details} \label{PO}

\subsection{Shape-based Tanimoto Score} \label{volume}
Following the molecular volume representation proposed in \citep{grant1995gaussian, grant1996fast}, each atom in the molecular structure is modeled as a spherical Gaussian function. An atom located at coordinates $R_i = (x_i, y_i, z_i)$ with a radius $\sigma_i$
is represented by a Gaussian density function: 
 \begin{equation*}
     \rho_i(r_i)=p_i\exp(-\alpha_ir_i^2) \quad with \quad\alpha_i= \frac{k_i}{\sigma_i^2},
 \end{equation*}
where $r_i = |r-R_i|$ is defined as a distance vector from the atomic center $R_i$, and parameters $p_i$, $k_i$ are atomic-type-specific chosen to ensure that the Gaussian atomic volume matches the van der Waals hard-sphere volume: $V_i^a=\frac{4\pi}{3}\sigma_i^3$, where the exponent $a$ refers to atom-based gaussian volumes.

According to the Gaussian product theorem, the product of two Gaussian spheres results in another Gaussian function. Utilizing this property, explicit formulas can be derived to compute the intersection volume between two atomic Gaussians, enabling the calculation of both molecular volumes and overlapping volumes between molecules \citep{grant1995gaussian, grant1996fast}. The intersection volume between atoms $i$ and $j$ is expressed as:
\begin{equation*}
    V_{ij}^a=p_ip_jK_{ij}(\frac{\pi}{\alpha_i+\alpha_j})^{3/2} \quad with \quad K_{ij} = \exp(-\frac{\alpha_i\alpha_jR_{ij}^2}{\alpha_i+\alpha_j}), 
\end{equation*}
where $R_{ij}$ is the distance between atoms $i$ and $j$.
The total volume of a molecule is expressed as:
\begin{equation*}
V^g=\sum_iV_i^a - \sum_{i<j}V_{ij}^a+\sum_{i<j<k}V_{ijk}^a- \: ... 
\end{equation*}
Similarly, the overlapping volume between molecules $A$ and $B$ is given by:
\begin{equation*}
    V_{AB}^g=\sum_{i\in A, j \in B}V_{ij}^a-\sum_{i,j\in A, k \in B}V_{ijk}^a-\sum_{i\in A, j,k \in B}V_{ijk}^a   +  \: ...
\end{equation*}
For computational tractability, we approximate the overlapping volume by truncating this series at second order, thus including only pairwise and triplet intersections.

\subsection{Pharmacophore scoring} \label{pharmaco}

\textbf{Pharmacophore-based Tanimoto Score}. Pharmacophores of a molecule are 3D arrangements of specific chemical features, such as hydrogen bond donors, hydrogen bond acceptors, hydrophobic groups, and aromatic rings. These features are identified using RDKit. Each pharmacophore feature is positioned at the barycenter of the corresponding atomic group and assigned a specific type (e.g., hydrogen donor, hydrogen acceptor, aromatic, etc.). Pharmacophores are modeled as spherical Gaussian functions with a fixed radius of $1.7$Å, comparable to the van der Waals radius of a carbon atom. The overlapping pharmacophore volume between molecules $A$ and $B$ is given by:
\begin{align*}
V_{AB}^p &= 
\sum_{i \in A,\, j \in B} V_{ij}^p \, \mathbb{1}_{\{T(i) = T(j)\}} \\
&\quad - \sum_{i, j \in A,\, k \in B} V_{ijk}^p \, \mathbb{1}_{\{T(i) = T(j) = T(k)\}} \\
&\quad - \sum_{i \in A,\, j, k \in B} V_{ijk}^p \, \mathbb{1}_{\{T(i) = T(j) = T(k)\}} \\
&\quad + \: \ldots
\end{align*}
where $V_{ij}^p$ represents the intersection volume between two Gaussians pharmacophores of types $T(i)$ and $T(j)$ from molecules $A$ and $B$. The indicator function $ \mathbb{1}_{\{T(i)= T(j)\}} $  ensures that only pharmacophores of the same type are considered for overlapping volume calculations. In practice, we approximated the overlapping pharmacophore volume by truncating this series at the first order, including only pairwise interactions.

\begin{figure}[h!]
  \centering
  \includegraphics[width=\linewidth]{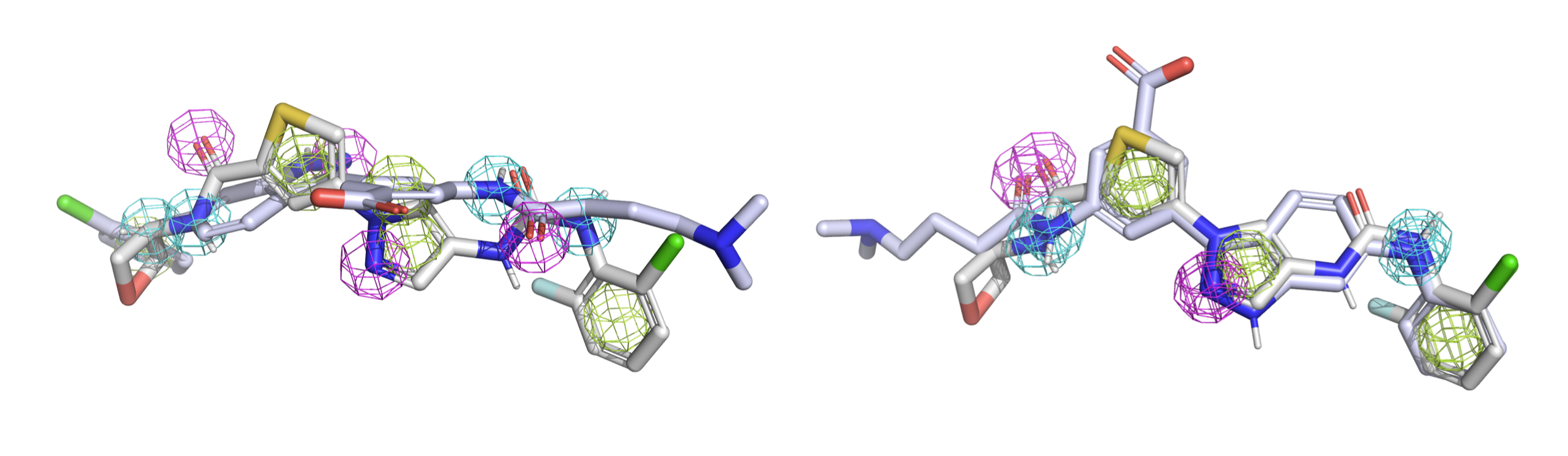}
  \caption{\textbf{Aligned ligands with pharmacophores overlapping}. Ligands carbons are shown in grey. Gaussian pharmacophore spheres are depicted as sphere mesh. The color of each sphere encodes a specific pharmacophoric feature: cyan for hydrogen bond donors, magenta for hydrogen bond acceptors, red for negatively charged groups, blue for positively charged groups, and limon for aromatic groups.}
  \label{fig:pharm}
\end{figure}

\textbf{Pharmacophoric complementarity}. The concept of pharmacophoric complementarity in our method stems from Gaussian-based shape similarity modeling. Specifically, once a Gaussian sphere is labeled with a pharmacophoric feature (e.g., hydrogen bond donor (HBD), hydrophobic, etc.), it becomes possible to compute a pharmacophoric similarity between two ligands based on the overlap of these labeled Gaussian volumes.

\begin{figure}[h!]
  \centering
  \includegraphics[width=0.85\linewidth]{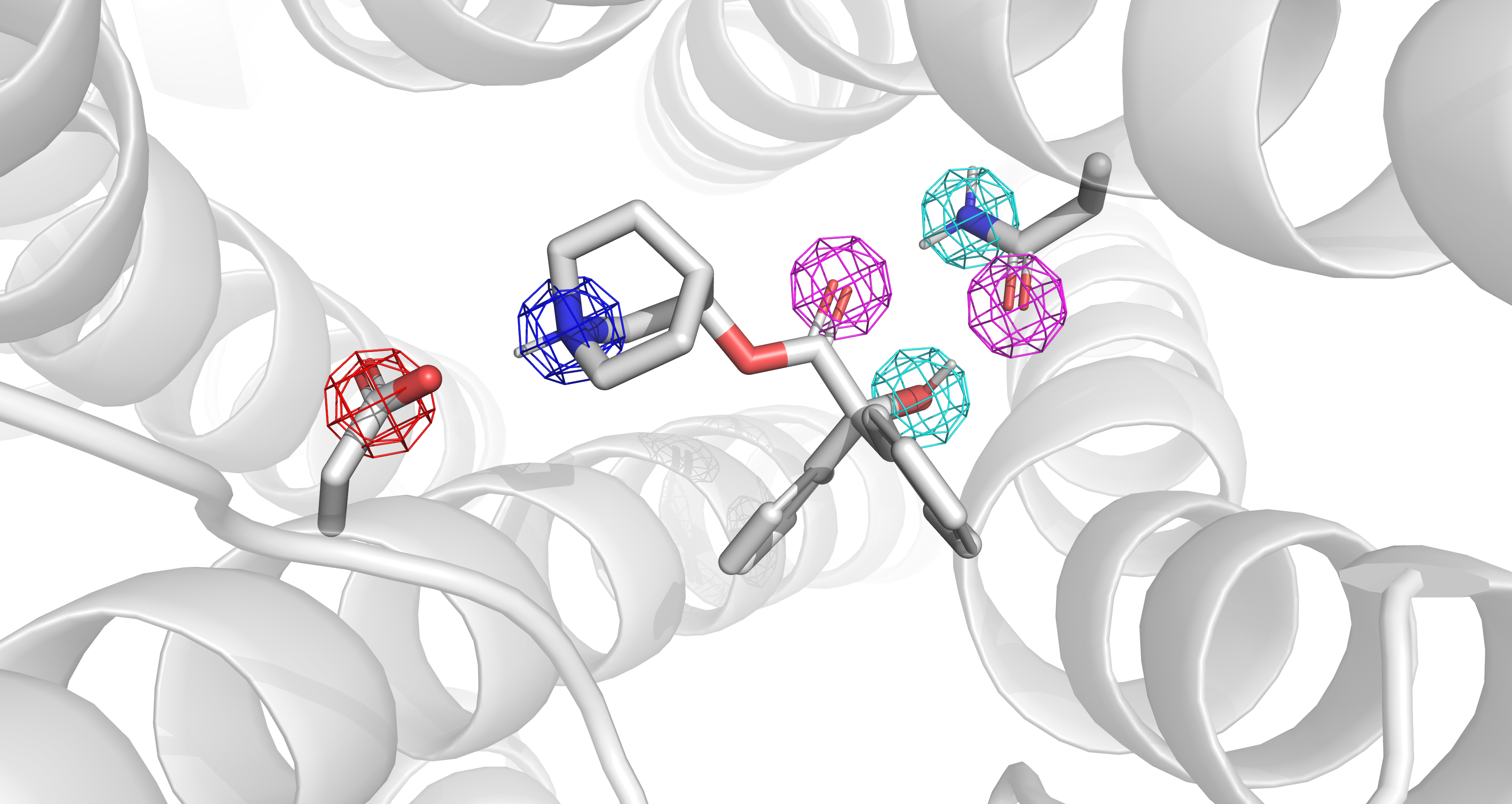}
  \caption{\textbf{Illustration of pharmacophoric complementarity within a \gls{PL} binding site}. 3D structure of the human M2 muscarinic receptor in complex with the ligand QNB (PDB ID: 5ZK3) \citep{suno2018structural}, shown in grey. Gaussian pharmacophore spheres are depicted as sphere mesh to illustrate key interaction features within the binding site. The color of each sphere encodes a specific pharmacophoric feature: cyan for hydrogen bond donors, magenta for hydrogen bond acceptors, red for negatively charged groups, and blue for positively charged groups.}
  \label{fig:pharmacophore_complementarity2}
\end{figure}

When introducing the protein pocket into the \gls{PO} process, we extended this idea to model complementarity between ligand and pocket by maximizing the overlap between complementary pharmacophoric Gaussians. For instance, \gls{HBD} and \gls{HBA} were defined as complementary, as well as cationic and anionic features (Figure \ref{fig:pharmacophore_complementarity2}). During the \gls{PO}, the $\mathcal{S}_{\mathrm{PTS}}$ term of the objective function increases the overlap between such complementary features from the ligand and the pocket, whereas the shape $-\mathcal{S}_{\mathrm{STS}}$ term prevents steric clashes by penalizing excessive overlap between atoms.

Regarding hydrogen bonding in particular, we model the \gls{HBD} pharmacophoric center as a Gaussian sphere located at the position of the donor hydrogen atom. If this hydrogen atom (and thus its associated Gaussian sphere) is positioned near an HBA-labeled Gaussian from the pocket, the overlap between these spheres is maximized, effectively favoring geometries consistent with hydrogen bond formation. The geometry naturally tends toward a favorable $\ce{X-H\cdot\cdot\cdot X}$ angle close to $180^{\circ}$, since this configuration maximizes Gaussian overlap. Simultaneously, the internal energy term in the PO loss encourages the donor group (e.g., $\ce{C-X-H}$) to remain in energetically favorable conformations.

\begin{figure}[h!]
  \centering
  \includegraphics[width=0.99\linewidth]{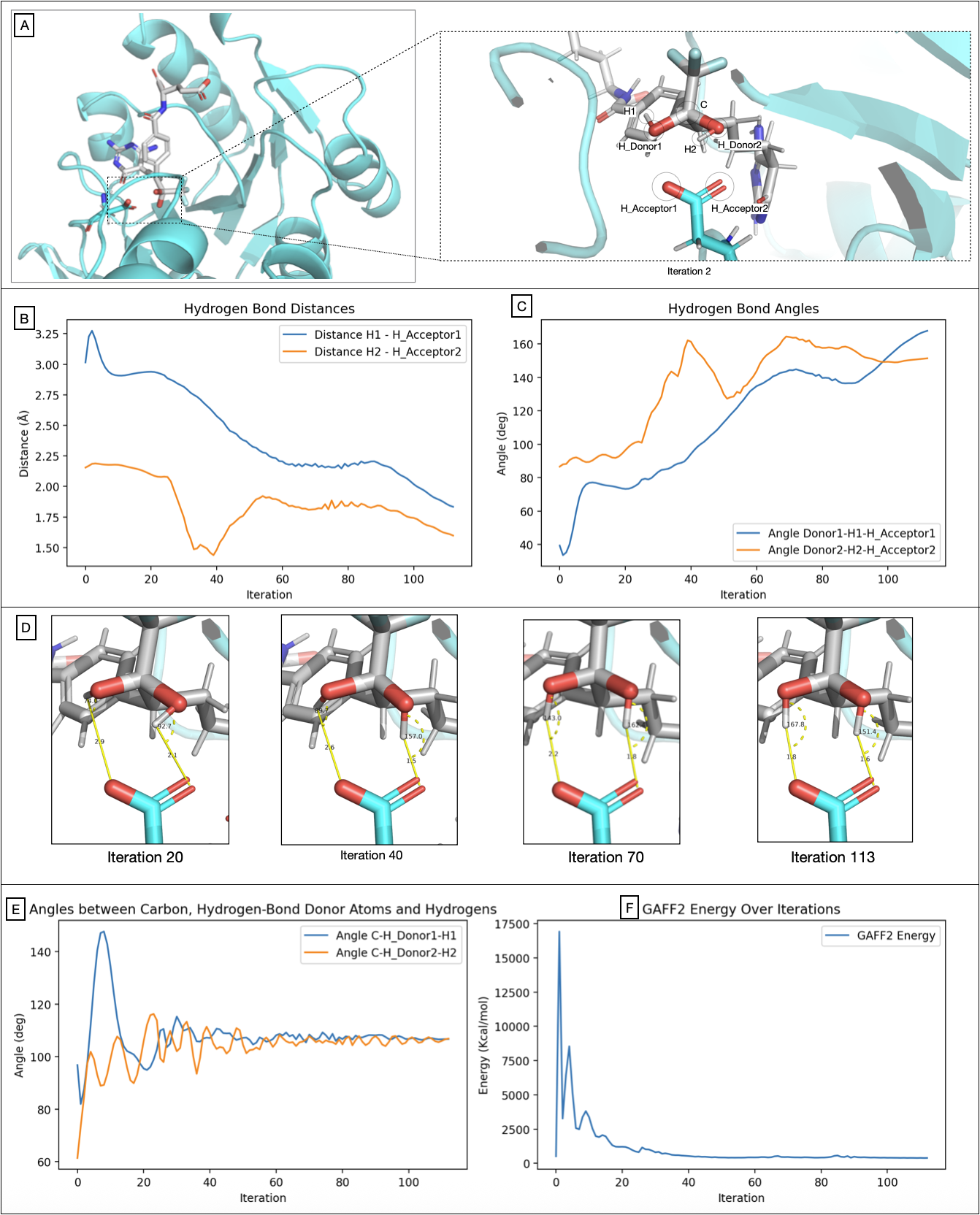}
    \caption{\textbf{Illustration of iterative optimization of hydrogen bonding interactions in a protein--ligand complex}. 
    3D structure of the human GAR Tfase (carbons in cyan) in complex with the hydrolyzed form of 10-trifluoroacetyl-5,10-dideaza-acyclic-5,6,7,8-tetrahydrofolic acid~\cite{zhang2003rational} (carbons in grey). 
    (A) The protein--ligand binding site shows two \gls{HBD} (Donor1, Donor2) and two \gls{HBA} (Acceptor1, Acceptor2). 
    (B) Time evolution of donor--acceptor distances ($\mathrm{H1\text{--}Acceptor1}$, $\mathrm{H2\text{--}Acceptor2}$) over the optimization iterations. 
    (C) Time evolution of hydrogen-bond angles ($\mathrm{Donor1\text{--}H1\text{--}Acceptor1}$, $\mathrm{Donor2\text{--}H2\text{--}Acceptor2}$). 
    (D) Representative ligand conformations at selected iterations (20, 40, 70, 113), illustrating the progressive stabilization of hydrogen bonds in the binding pocket. 
    (E) Evolution of the angles between the carbon atoms bound to hydrogen-bond donors and their respective hydrogens ($\mathrm{C\text{--}H_{Donor1}\text{--}H1}$, $\mathrm{C\text{--}H_{Donor2}\text{--}H2}$). 
    (F) GAFF2 energy profile across iterations, showing convergence from initially high-energy conformations toward a stabilized minimum.}
  \label{fig:hydrogen_bond_pharm_complementarity}
\end{figure}

As an illustrative example, consider the protein--ligand complex (PDB: \texttt{1NJS}) shown in Fig.~\ref{fig:hydrogen_bond_pharm_complementarity}. The ligand KEU (sequence number 510, chain A) contains two hydroxyl groups, \texttt{C5-OA1} and \texttt{C5-OA2}, both positioned near the carboxylate group of \texttt{ASP~144} (chain A):

\begin{itemize}
    \item The hydroxyl oxygen \texttt{OA1} belongs to a donor group (\texttt{C5-OA1-H1}) and lies near the acceptor \texttt{OD1} of \texttt{ASP~144}.
    \item The second hydroxyl oxygen \texttt{OA2} (in \texttt{C5-OA2-H2}) is close to the second acceptor \texttt{OD2}.
\end{itemize}

At the beginning of \gls{PO}, hydrogen atoms \texttt{H1} and \texttt{H2} are added by RDKit without considering the protein pocket. However, once optimization begins, complementary HBD Gaussians are assigned to these hydrogens, and their overlap with HBA Gaussians on \texttt{OD1} and \texttt{OD2} is maximized.

\subsection{Internal Energy components} \label{energy}
Internal energy components are computed using \gls{GAFF2} \citep{wang2004gaff2} force field parameters that depend on atomic and chemical bond properties. These parameters are parsed using the OpenMM\citep{eastman2023openmm} package and energy equations are implemented in PyTorch, ensuring differentiability with respect to ligand coordinates. Following OpenMM formulation, the internal energy is decomposed into two main components:
\begin{itemize}
    \item[-] \textbf{Bonded forces:} These include harmonic bond forces, harmonic angle forces, and periodic torsion forces, which account for bond stretching, angle bending, and torsional rotations.
    \item[-] \textbf{Non-bonded interactions:} These include Lennard-Jones interactions, modeling van der Waals forces, and Coulomb interactions, which account for electrostatic forces between partial atomic charges.
\end{itemize} 
Hence, internal energy is computed as follows: 
\begin{equation*}
    \mathcal{E}_{internal} = \mathcal{E}_{bond} + \mathcal{E}_{angle} + \mathcal{E}_{torsion} + \mathcal{E}_{lj} + \mathcal{E}_{coulomb}, 
\end{equation*}
where each term is described below.\\

\textbf{Harmonic Bond Energy} models the interaction between two bonded atoms $i,j$ using an harmonic potential:
\begin{equation*}
\mathcal{E}_{\text{bond}}^{ij} = \frac{1}{2} k (d_{ij} - d_0)^2,
\end{equation*}
where $d_{ij}$ is the bond length between atoms $i,j$,  $d_0$ is the equilibrium bond length, and $k$ is the bond force constant.

\textbf{Harmonic Angle Energy} describes the energy cost of deviating from the equilibrium bond angle between 3 atoms $i, j, k$:
\begin{equation*}
\mathcal{E}_{\text{angle}}^{ijk} = \frac{1}{2} k (\theta_{ijk} - \theta_0)^2,
\end{equation*}
where $\theta_{ijk}$ is the bond angle between $i,j,k$,  $\theta_0$ is the equilibrium bond angle, and $k$ is the angle force constant.

\textbf{Periodic Torsion Energy} accounts for dihedral interactions between four atoms $i,j,k,l$ using a periodic function:
\begin{equation*}
\mathcal{E}_{\text{torsion}}^{ijkl} = k (1 + \cos(n\phi_{ijkl} - \gamma)),
\end{equation*}
where $\phi_{ijkl}$ is the dihedral angle between $i,j,k,l$, $n$ is the periodicity, $\gamma$ is the phase offset, and $k$ is the torsional force constant.

\textbf{Lennard-Jones Energy} models van der Waals interactions between atoms $i,j$ using the Lennard-Jones potential:
\begin{equation*}
\mathcal{E}_{\text{lj}}^{ij} = 4\epsilon_{ij} \left[ \left(\frac{\sigma_{ij}}{d_{ij}}\right)^{12} - \left(\frac{\sigma_{ij}}{d_{ij}}\right)^{6} \right],
\end{equation*}
where $\epsilon_{ij}$ defines the interaction strength, $\sigma_{ij}$ is the distance at which the potential is zero, and $d_{ij}$ is the distance between atoms $i$ and $j$.

\textbf{Coulomb Energy} describes electrostatic interactions between atoms $i,j$ using Coulomb's law:
\begin{equation*}
\mathcal{E}_{\text{coulomb}}^{ij} = \frac{q_i q_j}{4\pi \varepsilon_0 d_{ij}},
\end{equation*}
where $q_i, q_j $ are the atomic charges, $d_{ij}$ is the distance between atoms $i$ and $j$, and $\varepsilon_0$ is 	the permittivity of vacuum ($\approx 8.85 \times 10^{-12} F/m$).

\subsection{Pose Optimization algorithm} 
The \gls{PO} algorithm is described in Algorithm \ref{algo:localoptim}. We used the Adam optimizer with a learning rate of 0.2. The refinement protocol is divided into two main phases, with weights that can vary across iterations to progressively adjust the optimization focus. The two-phase protocol is structured as follows:
\begin{itemize}
  \item \textbf{Phase 1: Hydrogen‐free refinement}
    \begin{itemize}
      \item \emph{Step 1}: Focus on minimizing internal energy.
      \item \emph{Step 2}:  Reinforce ligand alignment with a focus on shape and pharmacophore overlap.
      \item \emph{Step 3}: Add steric clashes and internal energy penalties while maintaining overlap improvements.
      \item \emph{Step 4}: Fine-tune steric clashes and internal energy, with reduced shape weighting.
    \end{itemize}
  \item \textbf{Phase 2: Hydrogen‐reintroduced refinement}
    \begin{itemize}
      \item \emph{Step 1}: Focus on internal energy minimization after hydrogen addition.
      \item \emph{Step 2}: Add shape and pharmacophore constraints for improved fit.
      \item \emph{Step 3}: Emphasize pocket volume interactions alongside energy adjustments to refine binding.
      \item \emph{Step 4}: Perform final fine-tuning of both overlap interactions and energetic stability.
    \end{itemize}
\end{itemize}

The weights of the loss terms (Table \ref{tab:optim_hyperparams}) were empirically tuned using a validation set of ten diverse protein–ligand complexes from the training data. For each term—shape similarity, pharmacophoric complementarity, steric clash penalty, and internal energy regularization—we explored multiple weightings. The final values were chosen based on convergence behavior, pose quality, RMSD and alignment performance across the validation set.

\begin{algorithm}[h!]
\caption{\textsc{PoseOptimization}}
\label{algo:localoptim}
\KwIn{Ligand conformation with coordinates $x$, number of optimization steps $n_{iterations}$, learning rate $lr$}
\For{$n_{iterations}$}{
    Compute loss: $\mathcal{L}_{\text{optim}} = \:  -
    \alpha \: \mathcal{S}_{\text{STS}} 
    - \beta \: \mathcal{S}_{\text{PTS}}
    - \omega \: \mathcal{S}_{\text{pocket}}+ \gamma \: \mathcal{E}_{\text{internal}}$\\
    
    Update atoms positions $x = x - lr \times \nabla_{x} \mathcal{L}_{optim}$
    } 
\end{algorithm}

\begin{table}[h!]
\centering
\caption{Hyperparameters for two‐phase pose optimization protocol. All weights were tuned empirically. $(x\rightarrow y)$ means that the weight increase linearly over iterations from $x$ to $y$.}
\vspace{0.7em}
\label{tab:optim_hyperparams}
\begin{tabular}{llrrrrr}
\toprule
\textbf{Phase} & \textbf{Step} & $\text{max\_iterations}$  & $\alpha$ & $\beta$ & $\omega$ & $\gamma$ \\
\midrule
\multirow{4}{*}{Without H} 
 & 1 & 50  & 1.0           & 0.0             & 0.0   & 1.0 \\
 & 2 & 200 & (1.0$\rightarrow$20.0) & (1.0$\rightarrow$50.0) & 10.0  & 10.0 \\
 & 3 & 50  & 30.0          & 10.0            & 200.0 & (1.0$\rightarrow$50.0) \\
 & 4 & 200 & 1.0           & 50.0            & 200.0 & 50.0 \\[6pt]
\multirow{4}{*}{With H} 
 & 1 & 25  & 0.0           & 0.0             & 0.0   & 1.0 \\
 & 2 & 200 & 10.0          & 10.0            & 25.0  & 10.0 \\
 & 3 & 50  & 10.0          & 10.0            & 150.0 & (10.0$\rightarrow$25.0) \\
 & 4 & 200 & 10.0          & 10.0            & 25.0  & 10.0 \\
\bottomrule
\end{tabular}
\end{table}

\section{AlignDockBench and training dataset}
\subsection{AlignDockBench}  \label{benchmark_set}
\begin{longtable}{lp{12.5cm}}
\caption{PDB IDs of template and query structures used in AlignDockBench.}\\
\label{tab:benchmark_set}\\
\toprule
\textbf{Template PDB ID} & \textbf{Query PDB IDs} \\
\toprule
1A4G & 2QWJ, 1INF, 1VCJ, 4HZW, 1INV, 2QWI, 1IVB, 1B9S, 1XOG, 3K37, 1F8E, 1F8C, 4MJV, 1INW, 5JYY \\ \midrule
2B1P & 7KSJ, 7KSI, 4W4V, 4WHZ \\ \midrule
1UY6 & 1UYD, 3HZ5, 1UYH, 5LR1, 1UY9, 7D24, 3FT8, 2H55, 7D22, 1UY8, 2FWZ, 6OLX, 1UYI, 4XIR, 1UYF, 4U93, 4XIQ, 6LR9, 1UY7, 1UYC, 7D26, 2FWY \\ \midrule
3BIZ & 3CR0, 2Z2W, 3BI6, 3CQE, 1X8B, 2IO6, 2IN6 \\ \midrule
2AA2 & 6W9M, 1ZUC, 6W9K, 5MWY, 4QL8, 5HCV, 1XQ3, 6W9L, 2A3I, 2OAX, 2AMB, 2Q1V, 1GS4 \\ \midrule
3QKL & 3QKM, 3QKK \\ \midrule
3EML & 5IUB, 5IUA, 5IU7, 5IU8 \\ \midrule
3SFF & 7ZZW, 3SFH, 7ZZU \\ \midrule
3D4Q & 3PPK, 3PSB, 3PPJ, 7SHV \\ \midrule
2ICA & 3M6F \\ \midrule
3LBK & 4OQ3, 5LAV, 1T4E \\ \midrule
3BQD & 7PRV, 6W9M, 6W9L, 5NFP \\ \midrule
3KBA & 3HQ5, 3G8O, 1ZUC \\ \midrule
2I0E & 8U37, 8UAK, 2JED \\ \midrule
1XOI & 2IEG, 2ZB2 \\ \midrule
3SKC & 4XV1, 4EHG, 5ITA, 4E4X, 5FD2, 3TV4, 4PP7 \\ \midrule
3KL6 & 2J34, 2VVU, 5VOF, 2P93, 2Y7Z, 2Y80, 2VVC, 2VWO, 4Y71, 1WU1, 2UWP, 2J38, 2UWO, 2J4I, 2P95, 1IOE, 1NFW, 1IQI, 2UWL, 2J2U, 1IQ,L 2VWL, 2VVV, 2J94, 1IQK, 2VWN, 1IQN, 2PR3, 4Y79, 1IQG, 2CJI, 4Y7A, 2J95, 4Y7B, 2D1J, 1J17, 4ZH8, 1IQH, 4Y76, 3SW2, 2P94 \\ \midrule
2XCH & 3OT8, 2R7B, 2IN6, 2PE1, 2IO6 \\ \midrule
1E3G & 3G0W, 1ZUC, 2NW4, 2AAX, 5T8J, 2HVC, 4QL8, 4ZN7, 8E1A 5V8Q 5T8E, 2IHQ, 2A3I, 2AMB, 5VO4, 5CJ6, 3D90, 1XNN, 1GS4 \\ \midrule
2E1W & 1WXZ, 1V7A, 1NDY \\ \midrule
3LAN & 3LAL, 1C1C, 2JLE, 1TL3, 3DRP, 3TAM, 3LAK, 4NCG, 2YNG, 8U6E, 1C1B, 2WON, 1RT1, 8U6P, 7SLR, 8U6C, 4WE1, 2B5J, 1RT2, 8U6B, 3DYA, 8U6O, 7SLS, 2BE2, 3LAM, 2YNI, 5TUQ, 8U6Q, 3T19 \\ \midrule
3LPB & 6VNI \\ \midrule
1UYG & 2QG2, 2QF6, 1UY9, 4CWR, 3HZ5, 8SSV, 4U93, 4R3M, 1UYI, 7D24 6N8Y 3B26 7D26 5LR1 7D22 1UYC 2H55 1UYK 3O0I, 1UY7, 1UYD, 3B25, 1UY8, 6LR9, 3FT8, 1UYM, 1UYH, 4L91, 6EL5, 2FWY \\ \midrule
3BKL & 3BKK, 6TT1 \\ \midrule
1W7X & 4JYU, 5PAM, 4X8V, 4ZXY, 1YGC, 1W0Y, 4YT7, 4JYV, 4YT6, 4NGA, 5PAQ, 4NG9, 1W2K \\ \midrule
1S3B & 2VZ2, 1GOS, 2C66, 2C64, 2C65, 1S3E, 5MRL \\ \midrule
3F9M & 4MLH, 3FR0, 6E0E, 3GOI, 3A0I, 4MLE \\ \midrule
2ZEC & 3V7T, 2ZA5, 2BM2, 2ZEB \\ \midrule
2GTK & 3FEJ \\ \midrule
3LN1 & 3N8X, 6COX, 5KIR, 3QMO \\ \midrule
2RGP & 3W33, 3BEL \\ \midrule
3CJO & 2FME, 2FL2, 6HKX, 2FKY, 2G1Q, 1YRS, 2FL6 \\ \midrule
3I4B & 5BVF 3I60 \\ \midrule
1Q4X & 3IMY \\ \midrule
2VWZ & 5MJA, 4YJS, 8BK0, 4YJP, 4P4C, 2VWY, 4G2F \\ \midrule
3HMM & 6B8Y, 3KCF, 1VJY, 1PY5, 3FAA, 5QTZ, 1RW8 \\ \midrule
5L7H & 2OAX, 5MWY, 5L7G \\ \midrule
2OI0 & 3B92 \\ \midrule
2AZR & 2BGE, 7FQU, 2HB1 \\ \midrule
3FRG & 5C2E, 5SHB, 4FCD, 5SF7, 3GWT, 6MSC, 5SHG, 5SJF, 5TKB, 4PM0, 5SKF, 3SNI \\ \midrule
1LRU & 1BSJ \\ \midrule
1ZW5 & 4OGU, 4DXJ, 3ICZ, 4DWB, 4KPD \\ \midrule
3L3M & 4RV6, 5XSR, 5A00, 5XST, 2RCW, 6I8M \\ \midrule
1EVE & 6TT0, 6EZH, 7D9Q, 6EZG, 7D9P, 7D9O, 5NAP \\ \midrule
3CHP & 7AV1 \\ \midrule
1F0R & 1F0S \\ \midrule
3NY8 & 2YCZ, 7DHI, 5A8E, 6PS1, 4BVN \\ \midrule
3KK6 & 3N8X, 5KIR \\ \midrule
3M2W & 3FYJ \\ \midrule
1L2S & 6DPY, 1XGJ, 1XGI, 4KZ4, 4JXS, 6DPX \\ \midrule
1B9V & 3CKZ, 1B9T 1VCJ, 3K39, 4DGR \\ \midrule
2QD9 & 6M95, 3ZSH \\ \midrule
3NW7 & 3NW6, 3I81 \\ \midrule
3K5E & 1X88, 2FL6, 2FME, 2X7C, 2X7D \\ \midrule
1J4H & 1J4I \\ \midrule
1S3V & 1MVT \\ \midrule
2ZDT & 2GMX, 2ZDU \\ \midrule
2VT4 & 6PS1, 5A8E, 2YCZ, 6PS4, 2Y04 \\ \midrule
1CX2 & 5KIR \\ \midrule
3KC3 & 3R30 \\ \midrule
1KVO & 1KQU, 1J1A \\ \midrule
\end{longtable}

\begin{figure}[h!]
  \centering
  \includegraphics[width=0.85\linewidth]{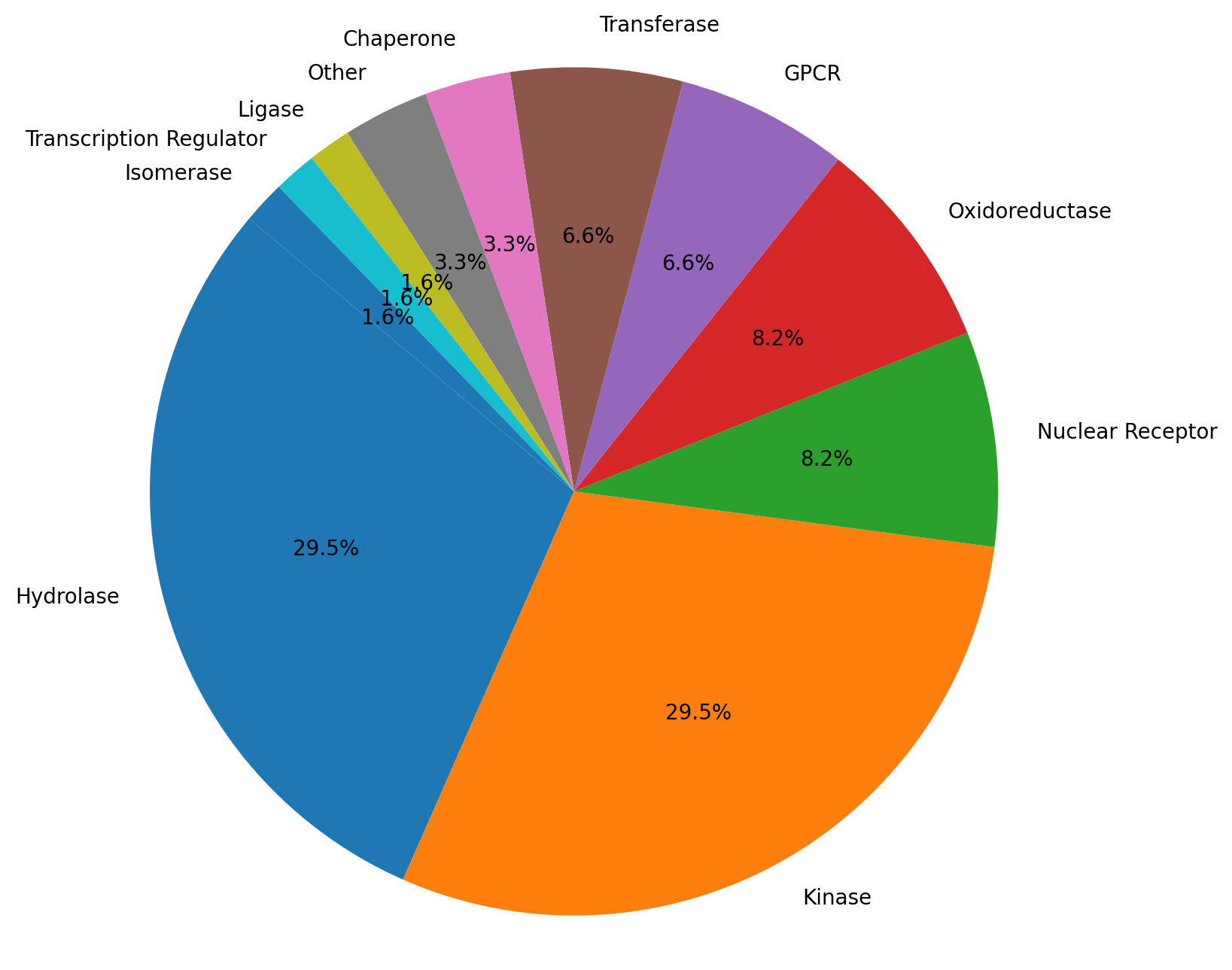}
  \caption{Overview of protein diversity in AlignDockBench.}
  \label{fig:benchmark_pruned_diversity}
\end{figure}

\begin{figure}[h!]
  \centering
  \includegraphics[width=0.85\linewidth]{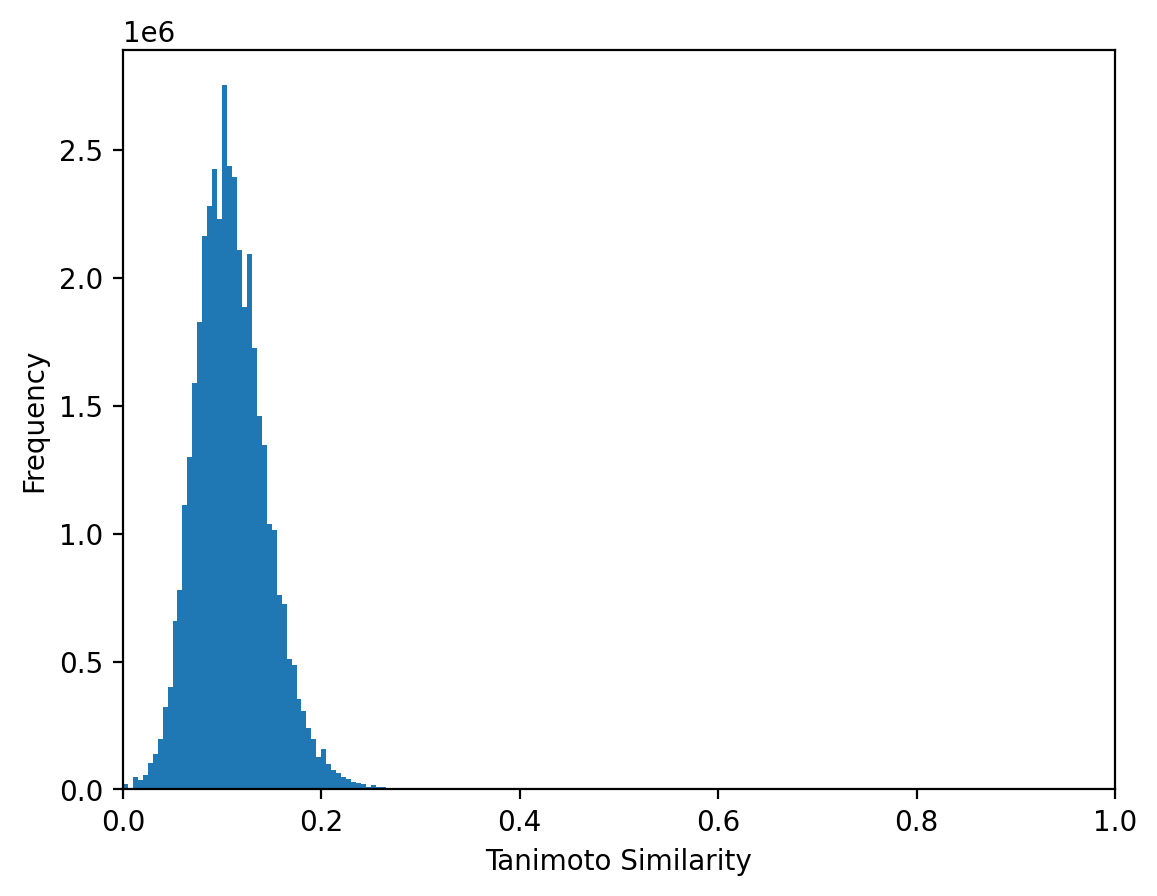}
  \caption{Morgan fingerprints tanimoto similarity histogram between AlignDockBench and the training set molecules.}
  \label{fig:hist_tan_sim_morgan_fp_plot}
\end{figure}

\subsection{Structure Preparation} \label{structure_prep}
Each structure retrieved from the \gls{PDB} was repaired and protonated using the PDB2PQR program \citep{dolinsky2007pdb2pqr} with the AMBER force field. Subsequently, energy minimization was performed using the GROMACS molecular dynamics engine \citep{van2005gromacs} with the AMBER03 force field in implicit solvent. Small molecules were parametrized using \gls{GAFF2} \citep{wang2004gaff2} with the ambertools suite \citep{case2023ambertools}, and partial charges were assigned using AM1-BCC \citep{am1bcc}. The minimization was performed in two stages: initially, 5,000 steps (time step = 2 fs) were conducted with positional restraints of 10,000 kJ/mol·nm² applied to all heavy atoms. In the second stage, another 5,000 steps were performed with restraints of 1,000 kJ/mol·nm² applied to the protein backbone and all heavy atoms of the small molecules.

\section{Experimental details} 
\subsection{Experimental setup} \label{experimental_set_up}

\textbf{Vina and rDock.} Docking simulations were performed as follows: a random conformer of each query molecule was generated using RDKit ETKDGv2 \citep{riniker2015better} and minimized with MMFF94 forcefields \citep{halgren1996merck}. Protonation states were assigned using the internal proprietary small-molecule protonation model at physiological pH (7.4). Ten docking poses were generated per molecule. In rDock, docking was carried out in "dock" mode using the default "STANDARD SCORE" scoring function. For AutoDock Vina, docking was performed with an exhaustiveness parameter set to 15, using the default Vina scoring function.

\textbf{FitDock and LSalign.} To prepare ligand inputs for both FitDock \citep{yang_fitdock_2022} and LS-align \citep{hu_ls-align_2018}, we generated 10 random conformers per query molecule using the same protocol described above, based on RDKit's ETKDGv2 method followed by MMFF94 energy minimization. Each minimized conformer was exported as an individual SDF file, then converted to MOL2 format using Open Babel \citep{o2011openbabel}, with partial atomic charges assigned using the Gasteiger method \citep{gasteiger1980iterative}. 
For each template–query pair, FitDock was provided with the MOL2 representations of both the template and query ligands, as well as the corresponding \gls{PDB} files of the template and query proteins. LSalign requires only the MOL2 ligand files; we used its flexible alignment mode with the “-rf” flag enabled, allowing torsions to adjust during superposition. After alignment, we selected the top-ranked conformer based on each software's primary scoring metric: the “Binding Score” for FitDock, and the “RMSD lb” output for LSalign. 

\textbf{ROSHAMBO.} For the ROSHAMBO software \citep{atwi_roshambo_2024}, the query ligand was provided as a single 2D SDF file, from which ten aligned poses were generated. Shape volumes were modeled using the "Gaussian" option. The resulting poses were ranked according to the “ComboTanimoto” score, and the highest-scoring pose was retained. 

\subsection{Supplementary results} 
\subsubsection{Cumulative distribution of RMSD values} 
\begin{figure}[h!]
  \centering
  \includegraphics[width=0.85\linewidth]{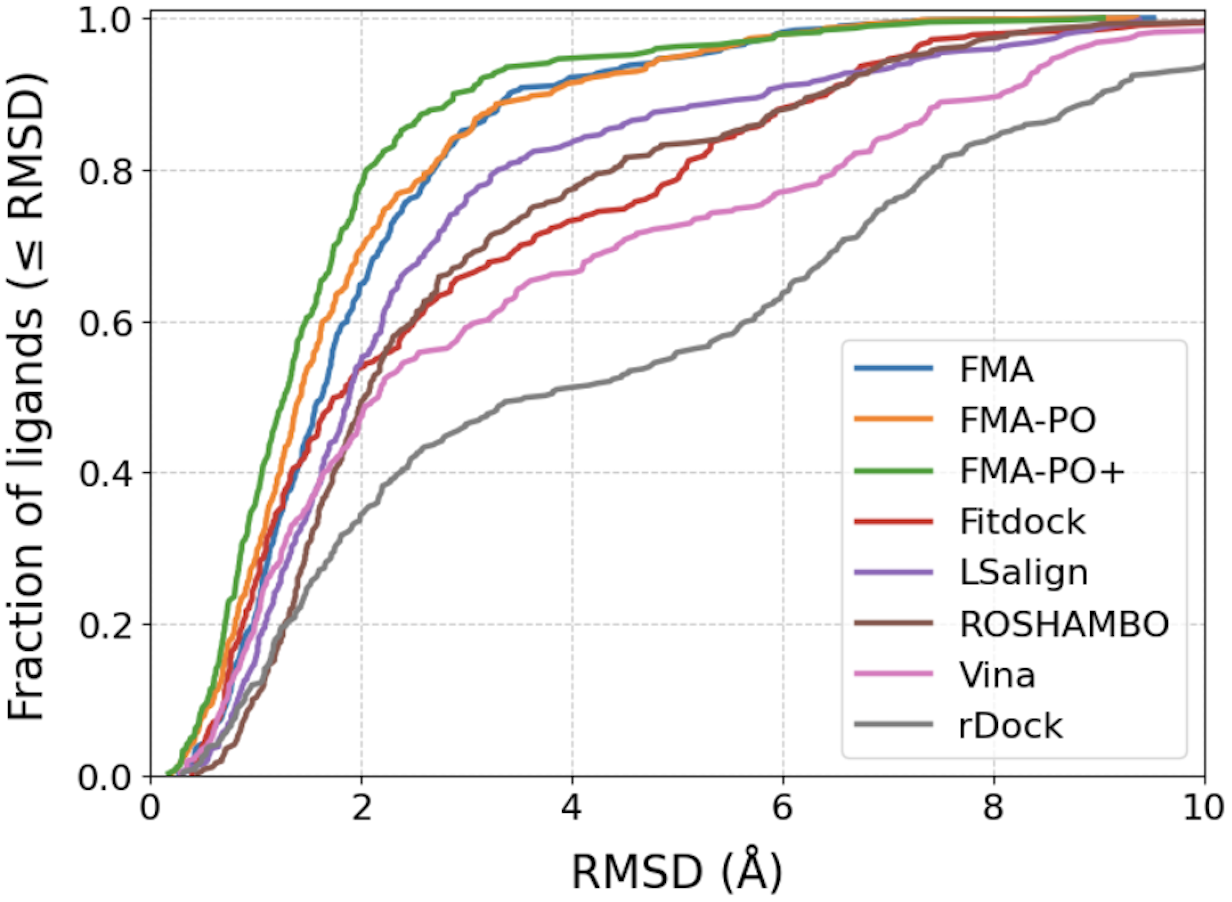}
  \caption{Cumulative distribution of RMSD values: percentage of molecules with RMSD below
different thresholds}
  \label{fig:cumulative_distrib}
\end{figure}

In Figure \ref{fig:cumulative_distrib} we report the cumulative distribution of RMSD values on AlignDockBench for each method.

\subsubsection{Redocking experiments} 
Results for each method in a redocking scenario, where the query ligand is aligned within its own crystal protein, are presented in Table \ref{tab:bench_results_redocking}. Methods like ROSHAMBO and LSalign, which do not incorporate protein information, have identical performance in both redocking and crossdocking contexts.

\begin{figure}[h]
  \centering
  \includegraphics[width=0.85\linewidth]{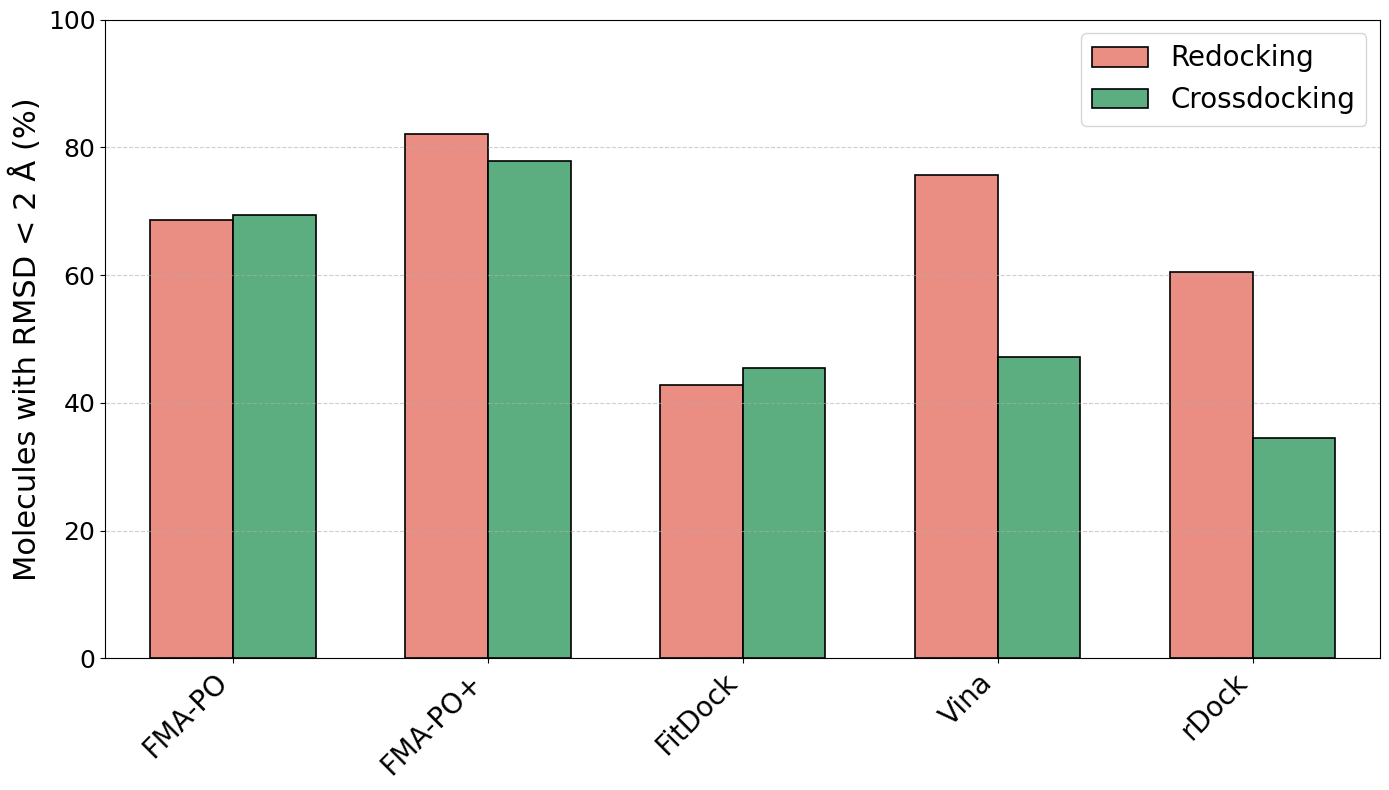}
  \caption{Pose accuracy comparison in redocking and crossdocking scenarios.}
  \label{fig:redocking_vs_crossdocking}
\end{figure}

Figure \ref{fig:redocking_vs_crossdocking} compares the performance of pocket-aware methods in both crossdocking and redocking scenarios, highlighting the gain of using true protein context for pose accuracy. Traditional docking approaches, such as Vina and rDock, exhibit significant improvements in redocking due to the availability of the correct pocket environment. \gls{FM-MAPO}+ also shows a slight performance gain in the redocking scenario whereas \gls{FM-MAPO} shows no notable difference.

\begin{table}
\caption{Performance comparison of 3D molecular alignment and docking methods on AlignDockBench in a redocking scenario. Methods marked with an (*) use GPU. For methods that did not align all 369 molecules, percentages are reported as X/Y, where the first value is calculated over aligned molecules only, and the second over the total set of 369 molecules.}
\label{tab:bench_results_redocking}
\vspace{0.7em} 
\begin{adjustbox}{max width=0.85\textwidth, center}
\renewcommand{\arraystretch}{1.1} 
\centering
\begin{tabular}{lcccc}
\toprule
\textbf{Method} & \makecell{\# of Molecules \\ Aligned} & \makecell{Mean RMSD \\ (Å ↓)} & \makecell{\% of Molecules \\ with RMSD \(<\) 2Å (↑)} \\
\midrule
FMA-PO$^*$   & 369/369 & 1.86 ± 1.43 & 68.56 \\
FMA-PO$+^*$   & 369/369 & \textbf{1.52 ± 1.24} & \textbf{82.11} \\
FitDock & 292/369 & 3.27 ± 4.64  & 54.11 / 42.82\\
LSalign & 368/369 & 2.54 ± 2.00 & 54.35 / 54.2  \\
ROSHAMBO$^*$ & 369/369 & 2.87 ± 2.09 & 48.51 \\
rDock & 369/369 & 2.79 ± 2.93  &  60.43  \\
Vina & 368/369 & 1.95 ± 2.54 & 75.82 / 75.61 \\
\bottomrule
\end{tabular}
\end{adjustbox}
\end{table}

\subsubsection{Comparison to HarmonicFlow} \label{hf}
We compared \gls{FM-MA} to the recent flow-based docking method HarmonicFlow \cite{stark2023harmonic} using the official  settings from the authors’ repository. As shown in Table \ref{tab:bench_harmonic_flow}, \gls{FM-MA} significantly outperforms HarmonicFlow in both mean \gls{RMSD} and success rate ($<2$Å), even when reporting HarmonicFlow’s best \gls{RMSD} across the 10 sampled poses. Applying \gls{PO} to HarmonicFlow improves results but still underperforms \gls{FM-MA}, highlighting the higher quality of \gls{FM-MA}’s pose distribution and the benefit of using the template ligand to guide docking.

\begin{table}[h!]
\caption{Comparison to Harmonic Flow}
\label{tab:bench_harmonic_flow}
\vspace{0.7em} 
\begin{adjustbox}{max width=\textwidth}
\renewcommand{\arraystretch}{1.1} 
\centering
\begin{tabular}{lccccc}
\toprule
\textbf{Method} & \makecell{\# of Molecules \\ with successful pose generation} & \makecell{Mean RMSD \\ (Å ↓)} & \makecell{\% of Molecules \\ with RMSD \(<\) 2Å (↑)}  \\
\midrule
FMA (Top score) & 369/369 & 1.97 ± 1.36 &  64.77 \\
FMA (Top RMSD) & 369/369 & 1.39 ± 0.85 &  84.55 \\
\midrule
FMA + PO$+$  (Top score) & 369/369 & 1.62 ± 1.33 & 77.78 \\
FMA + PO$+$  (Top RMSD) & 369/369 & 1.11 ± 0.85& 91.6 \\
\midrule
Harmonic-Flow (Top RMSD) & 356/369 & 4.47 ± 2.07 &   2.81/2.71 \\
Harmonic-Flow + PO $+$(Top RMSD) & 356/369 & 3.71 ± 2.22 &  19.94/19.24 \\
\bottomrule
\end{tabular}
\end{adjustbox}
\end{table}

\subsubsection{Discussion about runtime} \label{discussion_runtime}
We analyzed runtime as a function of ligand size for both FMA and PO. As shown in Table \ref{tab:time_analysis}, PO (2.7~s per pose, run on CPU) is slower than FMA (0.84~s for 10 poses, run on GPU). In both cases, runtime increases moderately with the number of heavy atoms and rotatable bonds (Pearson~$\sim$0.2), without signs of exponential growth.

\begin{table}[h!]
\caption{Time, * GPU}
\label{tab:time_analysis}
\vspace{0.7em} 
\begin{adjustbox}{max width=\textwidth}
\renewcommand{\arraystretch}{1.1} 
\centering
\begin{tabular}{lccc}
\toprule
\textbf{Method} & \makecell{Average \\ runtime (s)} & \makecell{Pearson correlation with  \\ the number of heavy atoms} & \makecell{Pearson correlation with  \\ the number of rotatable bonds} \\
\midrule
FMA$*$  & 0.84 s (for 10 poses)  &  0.20 & 0.23   \\
PO & 2.7 s (for 1 pose) & 0.19 & 0.23  \\
\bottomrule
\end{tabular}
\end{adjustbox}
\end{table}

Runtime is a critical factor for practical applications. Each molecule is processed independently, so scaling across multiple compute instances is straightforward. To illustrate this, we report in Table \ref{tab:runtime_cost_parallel_hours} the estimated runtime and cost of processing various dataset sizes using 100 AWS \texttt{g4dn.2xlarge} instances in parallel. This setup could be use within a de novo drug design pipeline, where screening around 50,000 molecules is a realistic and relevant scale. For ultra-large virtual screening campaigns (e.g., tens or hundreds of millions of molecules), further acceleration and simplification of PO would be required. For example, moving to a batched optimization strategy, rather than optimizing one molecule at a time, could significantly reduce total runtime.

\begin{table}[h!]
\caption{Estimated runtime (in hours) and cost (USD) using 100 AWS \texttt{g4dn.2xlarge} instances in parallel for different numbers of molecules.}
\vspace{0.7em} 
\centering
\begin{tabular}{lccc}
\toprule
\textbf{Number of molecules} & \textbf{50,000} & \boldmath{$10^5$} & \boldmath{$10^6$} \\
\midrule
Time FMA+PO (h)      & 0.49 & 0.98 & 9.83 \\
\midrule
Cost FMA+PO (\$)     & 41 & 82 & 826 \\
\bottomrule
\end{tabular}
\label{tab:runtime_cost_parallel_hours}
\end{table}

\subsection{Quality of predictions} \label{PoseBuster_check}
\begin{figure}[h!]
  \centering
  \includegraphics[width=0.9\linewidth]{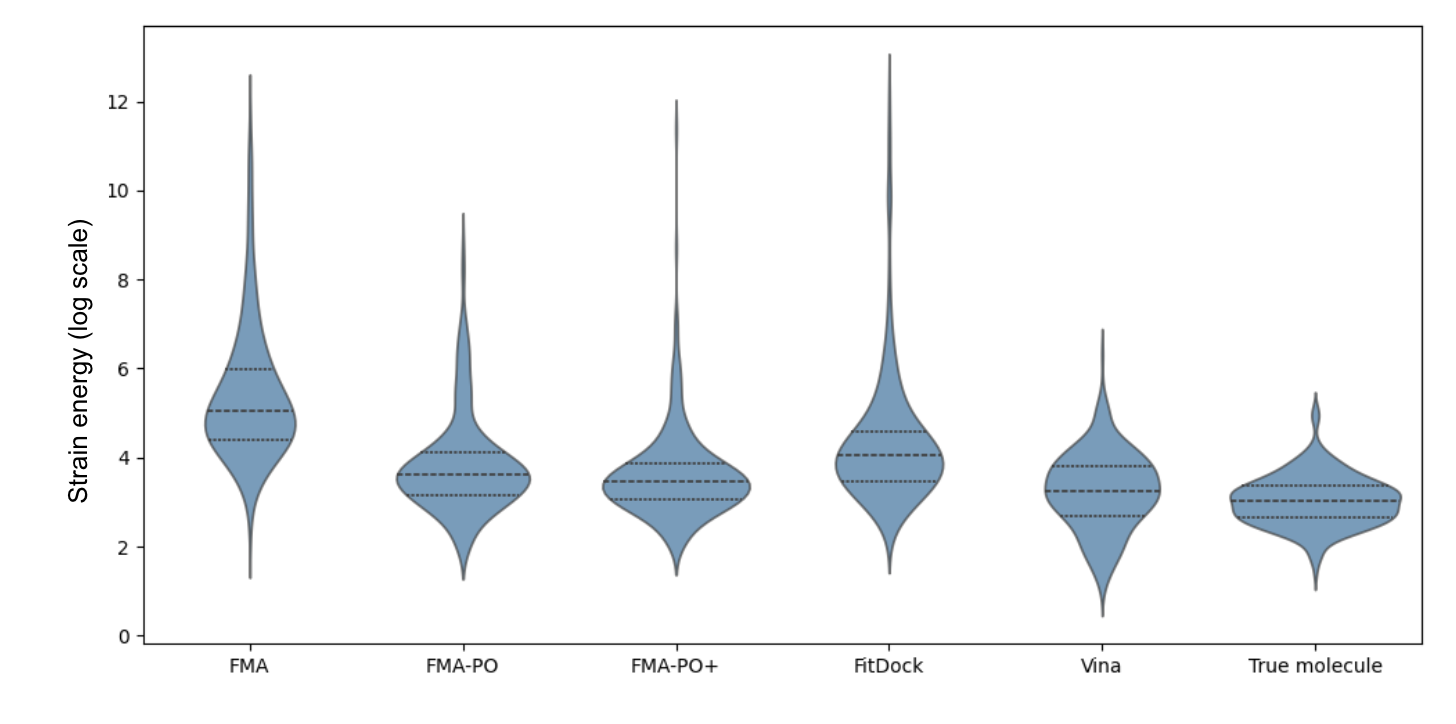}
  \caption{Strain energies of poses across methods (log scale).}
  \label{fig:strain_energies}
\end{figure}

Figure \ref{fig:strain_energies} reports of the strain energy as defined by GenBench3D \cite{baillif2024benchmarking}, for each method. FMA yields higher strain energies, due to the absence of the PO module. While unrealistic conformations are a known limitation of deep learning models \citep{buttenschoen2024posebusters}, this was one of the key motivations for introducing our PO module. Both FMA-PO and FMA-PO+ reduce strain energy.

Table \ref{tab:clash_counts} reports the number of molecules exhibiting clashes with protein, per method. FMA suffers from significantly more clashes, due to the absence of pocket conditioning. FMA-PO and FMA-PO+ reduce the number of clashes, confirming the effectiveness of the PO module in improving structural plausibility. 

To further evaluate the quality of generated poses, we employed the PoseBusters test suite \citep{buttenschoen2024posebusters}, which checks a range of geometric and chemical criteria to ensure physically plausible ligand conformations within the binding pocket. We observed that some of the $369$ ground-truth crystallographic query poses from AlignDockBench did not fully satisfy the PoseBusters criteria, particularly the minimum distance to the protein. As a result, we restricted our evaluation to a subset of 263 molecules for which the ground-truth poses passed all PoseBusters checks.

\begin{figure}[h!]
  \centering
  \includegraphics[width=0.8\linewidth]{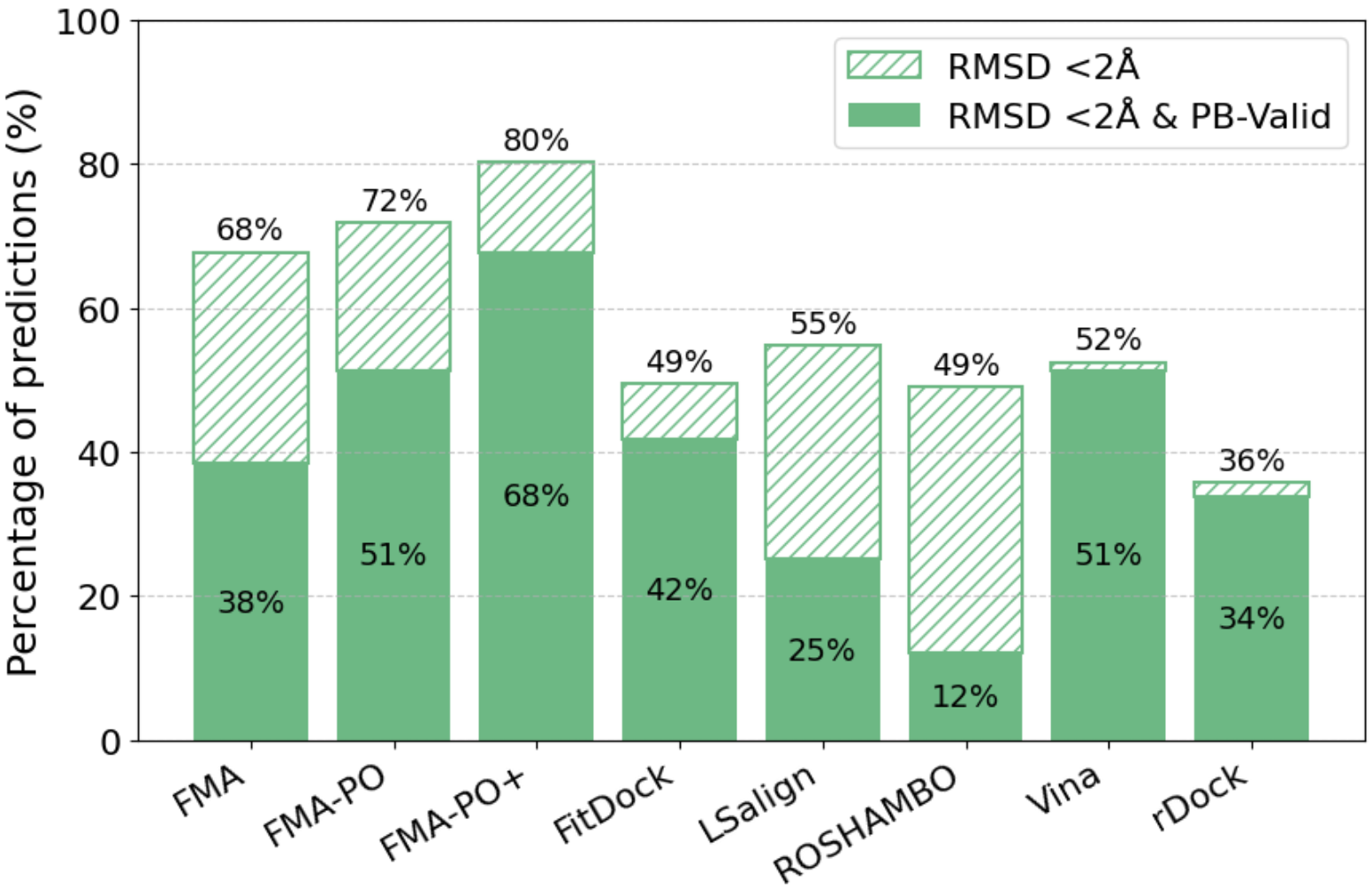}
  \caption{Percentage of accurate poses (RMSD $< 2$Å) and PoseBusters-valid (PB-valid) poses for each method, evaluated on a subset of 263 molecules whose ground-truth crystallographic poses pass the PoseBusters checks.}
  \label{fig:pb_valid}
\end{figure}

Figure \ref{fig:pb_valid} presents, for each method, the percentage of molecules achieving an RMSD below $2$Å, along with the proportion that also satisfies the PoseBusters criteria. Traditional docking methods tend to yield a higher proportion of PoseBusters-valid poses. In contrast, purely \gls{LB} methods like LSalign and ROSHAMBO exhibit a substantial drop in valid poses, reflecting the importance of protein context in guiding pose generation. \gls{FM-MA} (without \gls{PO}) shows a larger drop in PoseBusters-valid poses compared to \gls{FM-MAPO}+, highlighting the critical role of \gls{PO} and the importance of sampling multiple initial poses. Nevertheless, \gls{FM-MAPO}+ still exhibits a $12\%$ drop, primarily due to the minimum distance-to-protein check. This suggests that the method could be further improved by incorporating additional pocket-specific constraints or assigning greater weight to the pocket score during the \gls{PO} stage. Additionally, integrating protein context directly into the initial pose generation with the \gls{FM-MA} model may enhance the physical plausibility of the generated poses.

\begin{table}[h!]
\centering
\caption{Number of molecules with protein clashes with the protein, per method.}
\label{tab:clash_counts}
\begin{tabular}{lrrr}
\toprule
\textbf{Method} & \textbf{1 clash} & \textbf{2 clashes} & \textbf{$\geq$ 3 clashes} \\
\midrule
FMA                    & 37 & 16 & 20 \\
FMA-PO                 & 35 &  9 &  7 \\
FMA-PO+                & 21 &  6 &  2 \\
Vina                   &  1 &  0 &  0 \\
FitDock                &  7 &  5 &  4 \\
Co-crystallized Ligand &  0 &  0 &  0 \\
\bottomrule
\end{tabular}
\end{table}

\subsection{Ablation studies} \label{ablation_study}
\subsubsection{FMA's contribution} 
We generated 10 candidate poses per method (\gls{FM-MA}, RDKit‑centroid, Vina, LS‑align, FitDock‑LB), applied PO to each, and selected the top‑ranked pose. Results are reported in In Table \ref{tab:PO_applied_to_diff_base_alignments_1}. The RDKit‑centroid baseline places a random conformer at the binding‑site center. \gls{FM-MAPO}+ consistently yields the lowest RMSD among all alternatives, indicating that \gls{FM-MA} is the best initialization strategy for PO and that its accuracy is crucial.

\begin{table}[h!]
\caption{PO applied to different base alignments. All methods use flexible search mode and identical PO hyperparameters.}
\label{tab:PO_applied_to_diff_base_alignments_1}
\vspace{0.7em}
\centering
\begin{tabular}{lccc}
\toprule
\textbf{Method + PO} & \makecell{\# of Molecules \\ Aligned} & \makecell{Mean RMSD \\ (Å ↓)} & \makecell{\% of Molecules \\ with RMSD \(<\) 2Å (↑)}  \\
\midrule
FMA + PO$+$  & 369/369 & 1.62 ± 1.33 & 77.78 \\
rdkit + PO$+$  & 369/369 & 4.63 ± 1.94 & 10.33 \\
Vina + PO$+$  & 368/369 & 1.96 ± 1.87 & 71.47/71.27\\
LS-align + PO$+$  &357/369 &2.31 ± 2.07 & 63.31/61.25 \\
Fitdock LB + PO$+$ & 365/369& 2.78 ± 2.41& 52.6/52.03\\
\bottomrule
\end{tabular}
\end{table}

\subsubsection{\gls{PO}'s contribution} 
In Table \ref{tab:poAblation}, we evaluate the impact of the \gls{PO} stage and its objectives on pose prediction. \gls{FM-MA}, exhibits the poorest performance across all metrics, highlighting the importance of the pose refinement. Moreover, including the pocket during \gls{PO} enhances the quality of the generated poses, as evidenced by lower mean RMSD values and higher percentages of accurate alignments for both  \gls{FM-MAPO} and \gls{FM-MAPO}+.

\begin{table}[h!]
\caption{Ablation studies of PO loss terms on AlignDockBench in a crossdocking scenario.}
\label{tab:poAblation}
\vspace{0.7em} 
\begin{adjustbox}{max width=0.85\textwidth, center}
\renewcommand{\arraystretch}{1.1} 
\centering
\begin{tabular}{lccc}
\toprule
\textbf{Method} & \makecell{Mean RMSD \\ (Å ↓)} & \makecell{\% of Molecules \\ with RMSD \(<\) 2Å (↑)} \\
\midrule
FMA & 1.97 ± 1.36 &  64.77  \\
\midrule
FMA-PO & 1.86 ± 1.42 & 69.38  \\
FMA-PO w/o $\mathcal{S}_{STS}$  &1.98 ± 1.38 & 63.96\\
FMA-PO w/o $\mathcal{S}_{PTS}$  & 1.86 ± 1.38& 67.48\\
FMA-PO w/o  $\mathcal{S}_{\text{pocket}}$ & 1.88 ± 1.41 & 67.75  \\
FMA-PO w/o $\mathcal{E}_{\text{internal}}$ & 2.8 ± 0.91 & 13.82 \\
\midrule
FMA-PO$+$  & 1.62 ± 1.33 & 77.78 \\
FMA-PO$+$ w/o $\mathcal{S}_{STS}$  & 1.85 ± 1.32 & 66.67 \\
FMA-PO$+$ w/o $\mathcal{S}_{PTS}$ & 1.71 ± 1.33& 74.25\\
FMA-PO$+$ w/o $\mathcal{S}_{\text{pocket}}$ & 1.70 ± 1.39 & 75.61 \\
FMA-PO$+$ w/o $\mathcal{E}_{\text{internal}}$  & 2.98 ± 1.26 & 14.09 \\

\bottomrule
\end{tabular}
\end{adjustbox}
\end{table}

In Table \ref{tab:PO_applied_to_diff_base_alignments_2}) we applied \gls{PO} to top poses from other methods. \gls{PO} consistently reduced RMSD supporting its general effectiveness as a refinement stage. \gls{PO} also improves chemical plausibility and geometrical validity as demonstrated in Figure \ref{fig:strain_energies} and Table \ref{tab:clash_counts}

\begin{table}[h!]
\caption{Impact of PO on FMA, Vina, LS‐align and FitDock LB compared to the base aligners alone.}
\label{tab:PO_applied_to_diff_base_alignments_2}
\vspace{0.7em}
\centering
\begin{tabular}{lccc}
\toprule
\textbf{Method}           & \# Aligned & Mean RMSD (Å $\downarrow$) & \% RMSD $<$ 2Å ($\uparrow$)\\
\midrule
FMA                  & 369/369 & 1.97 ± 1.36 & 64.77  \\
FMA + PO  & 369/369 & 1.86 ± 1.42 & 69.38 \\
[4pt]
Vina                 & 368/369 & 3.39 ± 2.81 & 47.28  / 47.15  \\
Vina + PO & 368/369 & 3.24 ± 2.85  & 52.17 / 52.03  \\[4pt]
LS‐align             & 368/369 & 2.54 ± 2.00 & 54.35  / 54.20  \\
LS-align + PO &357/369 &2.47 ± 2.15 & 59.66/57.72 \\[4pt]
FitDock LB & 366/369 & 2.93 ± 2.36 & 49.73/ 49.32  \\
Fitdock LB + PO & 366/369& 2.75 ± 2.38& 52.88/52.3\\
\bottomrule
\end{tabular}
\end{table}

\subsubsection{Vina contribution to $\mathcal{S}_{\text{score}}$} 

We evaluated the influence of the $S_\text{Vina}$ component in the total score $\mathcal{S}_{\text{score}}$ by systematically ablating it across three key configurations: FMA, FMA-PO(i.e., scoring of FMA poses and PO applied to the top-ranked pose), and FMA-PO$+$(i.e., PO applied to all poses, and scoring of optimized poses). The results are summarized in Table \ref{tab:bench_Svina_contribution}. These results indicate that Vina scoring provides limited benefit when applied to unoptimized poses (FMA or FMA-PO), but can help select better candidates among refined poses (FMA-PO$+$). Our method still outperforms all baselines without using the Vina score in $\mathcal{S}_{\text{score}}$.

\begin{table}[h!]
\caption{Vina contribution to $\mathcal{S}_{\text{score}}$ }
\label{tab:bench_Svina_contribution}
\vspace{0.7em} 
\centering
\begin{tabular}{lccc}
\toprule
\textbf{Method} & \makecell{Vina term \\ in $ \mathcal{S}_{\text{score}}$} & \makecell{Mean RMSD \\ (Å ↓)} & \makecell{\% of Molecules \\ with RMSD \(<\) 2Å (↑)} \\
\midrule
FMA  & Yes & 1.97 ± 1.36 &  64.77  \\
FMA &No& 1.96 ± 1.35 & 64.77 \\
\midrule
FMA-PO &Yes & 1.86 ± 1.42 & 69.38 \\
FMA-PO &No&1.81 ± 1.40 & 69.38 \\
\midrule
FMA-PO$+$ &Yes &1.62 ± 1.33 & 77.78\\
FMA-PO$+$ &No& 1.68 ± 1.38 & 75.61 \\
\bottomrule
\end{tabular}
\end{table}

\subsubsection{Impact of the number of generated conformations} 
\begin{figure}[h!]
  \centering
  \includegraphics[width=0.8\linewidth]{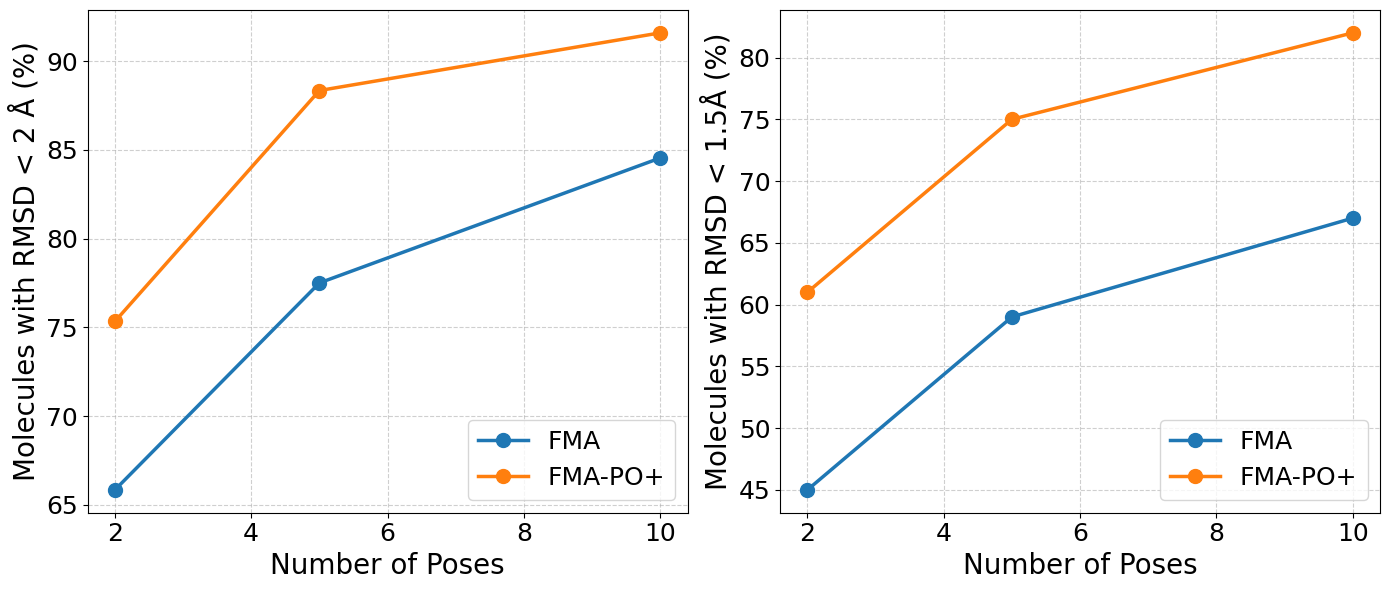}
  \caption{Effect of the number of generated poses prior to \gls{PO} on the pose accuracy of \gls{FM-MAPO} and \gls{FM-MAPO}+ on AlignDockBench.}
  \label{fig:comparison_number_samples}
\end{figure}

We assessed the impact of the number of generated poses with \gls{FM-MA} on final pose accuracy. Figure \ref{fig:comparison_number_samples} shows the percentage of molecules with \gls{RMSD} below $2$Å and $1.5$Å  for the best (i.e., lowest RMSD) pose of \gls{FM-MAPO} and \gls{FM-MAPO}+ as a function of the number of sampled poses prior to \gls{PO}. Performance increases with the number of generated poses, with higher sampling yielding a greater proportion of low \gls{RMSD} predictions.

\begin{figure}[h!]
  \centering
  \includegraphics[width=0.95\linewidth]{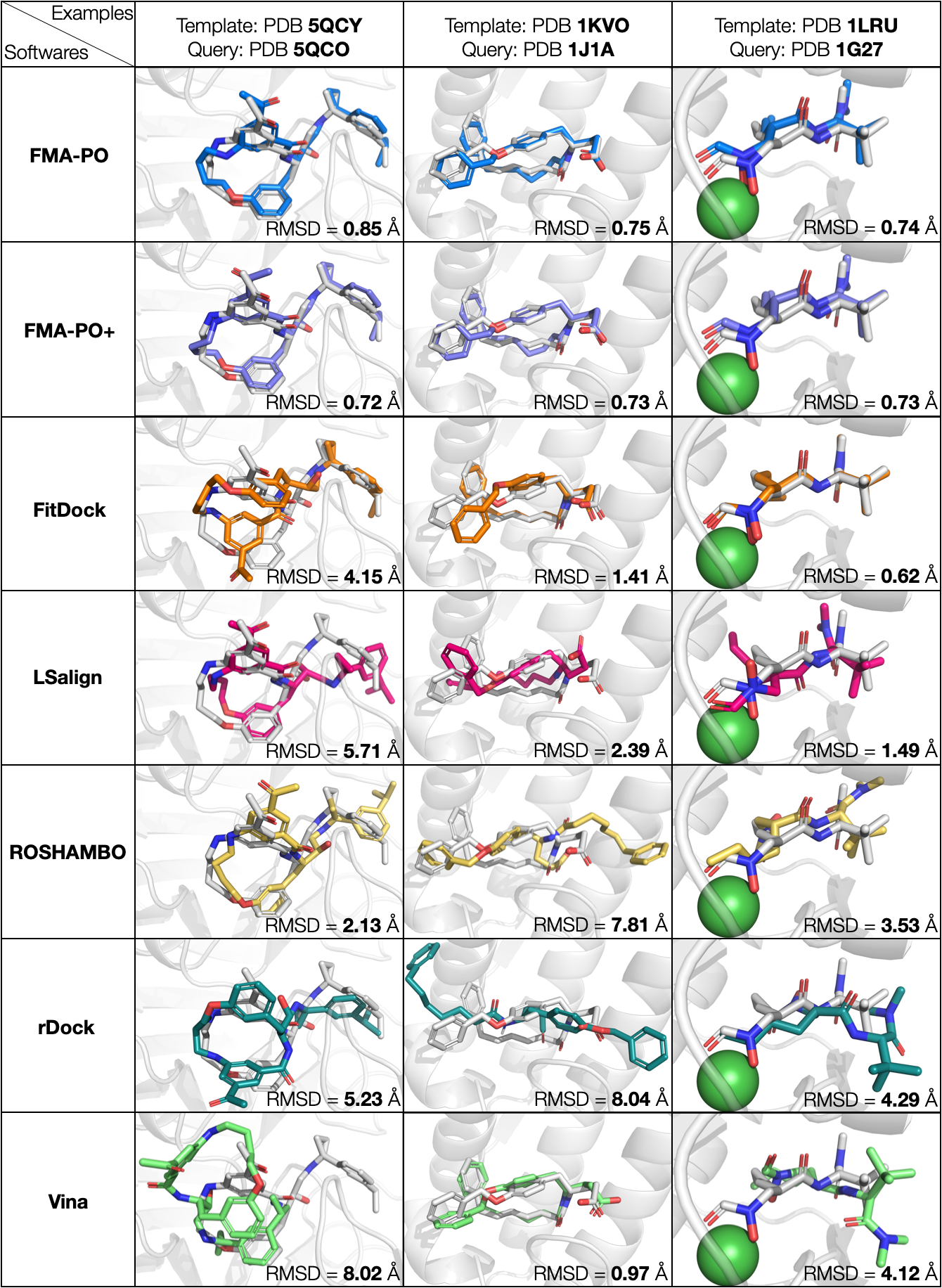}
  \caption{\textbf{Comparison of predicted binding poses of  various ligands using 3D molecular alignment and docking methods.} The crystallized protein and ligand are shown in grey. The first example is based on PDB entry 5QCO, corresponding to the human Beta-secretase 1 (BACE1), co-crystallized with a macrocyclic compound \citep{parks2020d3r}, using the PDB entry 5QCY \citep{parks2020d3r} as template. The second structure is based on PDB entry 1J1A, corresponding to the pancreatic secretory phospholipase A2 co-crystallized with an inhibitor \citep{hansford2003d}, using the PDB entry 1KVO \citep{cha1996high} as template. The third structure is based on PDB entry 1G27, corresponding to a polypeptide deformylase from \emph{Escherichia coli} co-crystallized with an inhibitor \citep{clements2001antibiotic}, using the PDB entry 1LRU \citep{guilloteau2002crystal} as template. The quality of the predicted poses was assessed by computing the \gls{RMSD} relative to the crystallized ligand conformation.}
  \label{fig:software_comparison_RMSDs}
\end{figure}

\end{document}